\documentclass[prc,aps,floats,floatfix,superscriptaddress,twocolumn,showpacs]{revtex4-1}
\usepackage{graphicx}
\usepackage{color}
\usepackage{amsmath}
\usepackage{amssymb}
\usepackage{booktabs}
\usepackage{mathtools}
\usepackage{multirow}
\usepackage{amsthm}
\usepackage{listings}
\usepackage{rotating}
\usepackage{booktabs}
\newcommand{\bra}[1]{\langle #1|}
\newcommand{\ket}[1]{|#1\rangle}

\begin{document}

%\title{\textit{Ab initio} many-body calculations of the $^4$He photo-absorption cross section}
\title{Operator evolution for {\em ab initio} electric dipole transitions of $^4$He} 

\author{Micah D. Schuster}
\email{mschuste@rohan.sdsu.edu}
\affiliation{San Diego State University, 5500 Campanile Drive, San Diego, CA 92182}
\author{Sofia Quaglioni}
\email{quaglioni1@llnl.gov}
\affiliation{Lawrence Livermore National Laboratory, P.O. Box 808, L-414, Livermore, CA 94551}
\author{Calvin W. Johnson}
\email{cjohnson@mail.sdsu.edu}
\affiliation{San Diego State University, 5500 Campanile Drive, San Diego, CA 92182}
\author{Eric D. Jurgenson}
\affiliation{Lawrence Livermore National Laboratory, P.O. Box 808, L-414, Livermore, CA 94551}
\author{Petr Navr\'{a}til}
\affiliation{TRIUMF, 4004 Wesbrook Mall, Vancouver, British Columbia, V6T 2A3 Canada}

\begin{abstract}
A goal of nuclear theory is to make quantitative predictions of low-energy nuclear observables starting from accurate microscopic %Hamiltonians. %
internucleon forces. 
A major element of such an effort is applying unitary transformations to soften the nuclear Hamiltonian  and hence accelerate the convergence of {\em ab initio} calculations as a function of the model space size. The consistent simultaneous transformation of external operators, however, has been overlooked in applications of the theory, particularly for nonscalar transitions. We study the evolution of the electric dipole operator in the framework of the similarity renormalization group method and apply the renormalized matrix elements to the calculation of the $^4$He total photoabsorption cross section and electric dipole polarizability. All observables are calculated within the {\em ab initio} no-core shell model. We find that, although seemingly small, the effects of evolved operators on the photoabsorption cross section are comparable in magnitude to the correction produced by including the chiral three-nucleon force and cannot be neglected.
\end{abstract}


\maketitle

\section{Introduction}

Unitary transformations of the Hamiltonian have been used to great effect in a range of nuclear physics problems \cite{okubo54,zheng93,stetcu07,nogga06,navratil00a,Barnea2000,Barnea2001,jurgenson11,roth11a,stetcu12,furnstahl13a}  to decouple high- and low-momentum components of the interaction and promote numerical convergence in large, but finite model spaces. 
However, in an $A$-nucleon system, such beneficial decoupling of momentum scales comes at the price of an effective Hamiltonian containing irreducible three- and higher-body (up to $A$-body) terms, even when initially absent. %starting from a bare two-nucleon ($NN$) force. %In addition, to obtain consistent expectation values and transitions matrix elements the same unitary transformation should be applied to any other operator besides the Hamiltonian. 
In addition, for consistency the same unitary transformation must to be applied to any operator associated with measurable quantities. This, once again, will induce many-body operators. 

Widely adopted is the similarity renormalization group (SRG) method, which employs a continuous unitary transformation %that can be 
of the Hamiltonian 
characterized by a momentum resolution scale $\lambda$~\cite{bogner07}.   
The SRG transformation (or, evolution) of the Hamiltonian has been carried out up to the three-body level both on a harmonic oscillator (HO) basis~\cite{jurgenson09a,jurgenson11,roth12a,Dicaire2014}  and, more recently, in momentum representation~\cite{Wendt2013}, and the resulting interactions have been successfully applied to compute properties of a variety of nuclei~\cite{furnstahl13a,jurgenson09a,jurgenson11,roth11a,tsukiyamat11a,roth12a,hergert13a,hupin2014}. %(\%expand or change citations here\%) %working within bound-state techniques.
     
%In this approach, as long as the forces in the initial Hamiltonian follows a hierarchy of decreasing strength (i.e., the contribution of three-nucleon ($NNN$) force is smaller than that of the two-nucleon($NN$) interaction, etc.), so do the induced many-body terms~\cite{jurgenson11}. 
%Further, %previous work suggest that, 
For %the low-lying energy spectrum of 
systems with up to $A\simeq 10$ nucleons, %truncating the
bound-state calculations including 
up to three-body 
induced forces 
%at the three-body level has been shown 
%up to the three-body level 
have been shown to lead to energies  mostly independent of $\lambda$ above $1.8$ fm$^{-1}$, i.e.\ %stopping at three-body forces 
%it is sufficient 
to approximately preserve the unitarity of the transformation~\cite{jurgenson09a,jurgenson11,roth12a}.  
%In particular, SRG resolution scales %above $\lambda\sim1.8$ fm$^{-1}$ 
%around $\lambda\simeq2$ fm$^{-1}$ have been found to provide a good balance between improved convergence and growth of induced higher-body terms. % for energy calculations \cite{jurgenson09a,jurgenson11,roth12a} %(\% Need more refs here here we will have to cite also the work of Roth et al.).
Small variations of the SRG momentum scale around 2 fm$^{-1}$ have been also shown to produce mostly negligible differences in  %More limited studies with respect to the range of adopted $\lambda$ values have been also performed for 
$n$-$^4$He~\cite{Hupin2013} and $n$-$^8$Be~\cite{Langhammer2015} elastic phase shifts, but a more quantitative investigation was not possible due to a slower rate of convergence for larger $\lambda$ values combined with the high computational demand.
%Because of this, the SRG approach has been successful when computing energies for a variety of nuclei.(\% Here we need to include the citations we had in the rapid report, plus other if needed\%)
%We can probably do without this

%There is more to nuclei than energy spectra, however. For instance, we want to accurately quantify 
%electric dipole transitions which lead to important observables that are difficult to measure such as, 
%e.g., the polarization of a nucleus \cite{stetcu09a} (\%add citation of new paper of Sonia\%); or the radiative capture $^7$Be$(p, \gamma)^8$B, 
%crucial to understanding the neutrino signature of our sun \cite{adelberger11a,navratil11a}. 
%The consistent evolution and application of operators are essential steps to enable an accurate description of these and other measurable nuclear properties within the SRG approach.

Few studies have dealt with %An 
the consistent transformation and application of operators,  the other component required for an accurate description of measurable nuclear properties when using effective interactions. %within the SRG approach. %is the consistent evolution of operators.
%Much fewer studies have been dealing with the consistent evolution and application of operators.  
%The SRG evolution of operators 
This was first studied using the Okubo-Lee-Suzuki (LS) renormalization \cite{okubo54,suzuki80,suzuki82} to compute electromagnetic properties for several nuclei \cite{stetcu05}.
For the SRG, the evolution of operators was achieved for the first time in the deuteron, where only one- or two-body operators are relevant, working in a momentum representation~\cite{anderson10}.  The more complicated process of evolving and applying operators in finite nuclei beyond the deuteron was first examined in Ref.~\cite{schuster14a}.  There, working on a translationally invariant HO basis, we extended the approach of Ref.~\cite{jurgenson09a} to evolve scalar (i.e., rank-zero in both angular momentum and isospin) operators in the two- and three-body spaces and used the resulting matrix elements to calculate expectation values on the ground state (g.s.)\  %the root-mean-square radius and total electric dipole strength 
of the $^4$He nucleus. (Note that only scalar operators contribute to expectation values for this $J^\pi T=0^+0$ four-nucleon state). In particular, %the root-mean square radius and the total electric dipole strength, 
we showed that the inclusion of up to three-body matrix elements in the
$^4$He nucleus all but completely restores the invariance of the %expectation values 
root-mean square radius and total electric dipole strength
under the SRG transformation.  %However, a more general approach capable to address operators inducing transitions between states of different total angular momentum, parity and/or isospin is still missing.

%However, for this first study we limited ourselves to scalar operators. This simplified considerably the task, but allowed onl 

While the work of Ref.~\cite{schuster14a} allowed us to perform initial proof-of-principle calculations, a general description of observables also requires the ability to evolve, and embed in finite nuclei, nonscalar operators. %as well as an accurate study  an accurate  
Further, %only a few studies have explored the dependence of scattering phase shifts on the SRG momentum scale $\lambda$ and 
more work is needed to accurately asses the consistency of the SRG approach for the description of continuum observables. % in the continuum of energies.   
Starting from an initial nucleon-nucleon plus three-nucleon ($NN+3N$) Hamiltonian from chiral effective field theory \cite{epelbaum09a,machleidt11a}, in this paper we present the first application of the SRG approach to compute the $^4$He photoabsorption cross section and electric dipole polarizability.  All induced forces up to the three-body level are retained in the transformed Hamiltonian, while the leading electric dipole transition operator is determined (for the first time) by evolution in the $A=2$ system. All calculations are performed within the {\em ab initio} no-core shell model (NCSM) \cite{navratil00b} working with translationally invariant harmonic oscillator (HO) basis states. The photoabsorption cross section is computed by means of the  Lorentz integral transform (LIT) method~\cite{efros94a,efros07a}, while the electric polarizability is obtained according to Podolsky's technique~\cite{Podolsky1928}. %which allows the ground state to be used as the driving term for the Lanczos- moment method [ ], which is our numerical method of choice for solving the Schr\"odinger equation. 
This %trick 
allows us to bypass the direct calculation of scattering states and to work only with square-integrable basis states. 

An {\em ab initio} investigation of both the photoabsorption cross section~\cite{quaglioni07}  and the electric polarizability~\cite{stetcu09a} of the $^4$He nucleus based on chiral $NN+3N$ interactions had already been accomplished in the past using LS effective interactions at the three-body cluster level~\cite{Navratil2002,navratil03a}, albeit  without renormalization of the electric dipole operator.     %obtained with the Lee-Suzuki unitary transformation \cite{suzuki80,suzuki82}.
The primary purpose of the present work is to %test the performance and consistency of the SRG approach 
use these observables as testing grounds to explore the performance and consistency of the SRG approach. In particular we will perform the first accurate investigation of the dependence on the SRG momentum scale of a continuum observable within a large range of $\lambda$ values.

The paper is organized a follows. 
Sec. II provides background on the %approach we use to compute the total photo-absorption cross section
formalism adopted. In particular, we discuss how the SRG method modifies the Hamiltonian and external operators and how the LIT can be used to compute the response induced by the an external perturbation, in our case, the dipole operator. In Sec. III we describe our results in three parts: convergence of the observables computed with respect to the size of the NCSM model space adopted, a discussion on the unitarity of the SRG transformation in our context and a comparison to experimental cross section data. Lastly, Sec. IV gives a brief summary of our results and describes the next steps in this research.

\section{Background}

%\subsection{Hamiltonian and many-body solution method}
\subsection{Hamiltonian and spectral resolution method}

%The increased capacity of supercomputers has stimulated the development of new algorithms to solve the nuclear many-body problem. The \textit{ab initio} no-core shell-model (NCSM) \cite{navratil00a} is one such method in which all nucleons are taken to be active, a practice generally restricted to few-body systems because the number of many-nucleon states quickly becomes unmanageable. 

%We describe nuclei as systems of $A$ non relativistic pointlike nucleons (protons and neutrons) interacting through free-space two- ($NN$) and three-nucleon ($3N$) forces 
We start with the intrinsic nonrelativistic Hamiltonian for a system of $A$ nucleons (protons and neutrons) %is
\begin{equation}
\hat{\text{H}}=\frac{1}{A}\sum_{i<j}\frac{(\vec{p}_i-\vec{p}_j)^2}{2M_N}+\sum_{i>j}^AV_{ij}^{NN}+\sum_{i>j>k}^AV_{ijk}^{3N}\,, %+\dots,
\label{eq:ham}
\end{equation}
where $V_{ij}^{NN}$ and $V_{ijk}^{3N}$ are, respectively,  two- and three-nucleon free-space interactions, which depend on the relative coordinates (and/or momenta for nonlocal forces) between particles, $\vec{p}_i$ is the momentum of particle $i$, and $M_N$ is the nucleon mass. 
We then look for the eigenfunctions of  $\hat{\text{H}}$ in the form of expansions over a complete set of translationally invariant and fully antisymmetric $A$-body states. This amounts to diagonalizing the Hamiltonian in the many-body basis.   %One then diagonalizes this Hamiltonian in a many-body basis. 
In particular, we use %a translationally invariant 
the Jacobi-coordinate harmonic oscillator (HO) basis of the {\em ab initio} (NCSM)~\cite{navratil00b}, %which depends only on Jacobi relative coordinates. %, see Appendix \ref{sec:jac}.
in which the model space is defined by all $A$-body states up to a maximum excitation of $N_{\rm max}\hbar\Omega$ above the minimum energy configuration of the system, and $\Omega$ is the HO frequeny.  
%This has the unique advantage to allow for easy separation of center of mass motion from the internal degrees of freedom \cite{navratil00a}. 
%When necessary, in the reminder of this paper, we will refer to the size of the HO spaces utilized to describe the $A=2$ and $A=3$ system as $N_{A2\rm{max}}$ and $N_{A3\rm{max}}$, respectively.

While in principle the above is an exact prescription for the solution of the Schr\"odinger equation associated with the Hamiltonian of Eq.~(\ref{eq:ham}), in practice we %truncate the size of the HO basis at 
work with a finite model space %value ofcharacterized by the parameter 
%of $N_\text{max}$ and aim at reaching convergence converging to the exact results with the increasing of this parameter.
and achieve convergence to the exact results with increasing $N_{\rm max}$. Crucial for the success of this approach is the use of unitary
transformations of the Hamiltonian chosen to reduce the coupling between high- and low-momentum states, which arises from the strong short-range repulsion of the bare nuclear interaction and leads to slow convergence in the
size of the model space.  Here we focus on the unitary transformation described by the SRG approach, outlined in the next section. 

Our numerical method of choice for obtaining the spectrum of energy states of the Hamiltonian is the Lanczos method \cite{lanczos}. Given a starting arbitrary unit vector $|\phi_0\rangle$, it recursively allows us to define a set of orthonormal basis states $|\phi_i\rangle$ -- known as Lanczos vectors -- for which the Hamiltonian matrix assumes a tridiagonal form:
\begin{align}
	b_{i+1}|\phi_{i+1}\rangle = \hat{\text{H}}|\phi_i\rangle - a_i|\phi_i\rangle -b_{i}|\phi_{i-1}\rangle\,.
\end{align}
Here $|\phi_{-1}\rangle = 0$, and $a_i = \langle\phi_i| \hat{\text{H}}|\phi_i\rangle$ and $b_{i} =\| b_{i}|\phi_{i}\rangle\|$ %$b_{i} = \langle\phi_{i-1}| \hat{\text{H}}|\phi_{i}\rangle$ 
are respectively the diagonal and upper (lower) diagonal elements of the Hamiltonian in the new basis, or  Lanczos coefficients as they are often called. The power of the Lanczos method is that the extremum eigenvalues of the Hamiltonian quickly converge to their true value after a limited number of  iterations, much smaller than the dimension of the problem. Further, relevant to the calculation of the $^4$He photoabsorption cross section and electric polarizability discussed in this paper, the Lanczos coefficients can be used to accurately evaluate the expectation value of the Green's function on a normalized vector,  $G(z)= \langle\phi_0|(z-\hat{\text{H}})^{-1}|\phi_0\rangle$, in terms of the continued fraction \cite{haydock74,haxton05a}
\begin{align}
%	G(z) &= \left\langle\phi_0\left| \frac{1}{z-\hat{\text{H}}} \right|\phi_0\right\rangle\nonumber\\
	G(z) = & \frac{1}{z-a_0-\frac{b^2_1}{z-a_1-\frac{b_2^2}{z-a_2-\frac{b_3^2}{\phantom{z}\ddots}}}}\,.
	\label{eq:gf}
\end{align}

\subsection{SRG evolution}
\label{sec:srg}
%We start by renormalizing the Hamiltonian via the SRG method. 
As implemented for nuclear physics \cite{bogner07,bogner10}, the SRG method employes %this is a series of 
a %continuous
unitary transformation, $U_s$, on the initial Hamiltonian $\hat{H}_{s=0} = \hat{\text{H}} $
\begin{equation}
\hat{H}_s=\hat{U}_s\hat{H}_{s=0}\hat{U}_s^\dagger,
\end{equation}
%depending on the continuos parameter $s$. %labels the sequence of Hamiltonians. 
%This transformation 
that can be implemented as a flow equation~\cite{wegner94} in the continuous parameter $s$ and an anti-Hermitian generator $\hat{\eta}_s=(d\hat{U}_s/ds) ~\hat{U}_s^\dagger$,
\begin{equation}
%\frac{d\hat{H}_s}{ds}=[ [\hat{T},\hat{H}_s],\hat{H}_s].
\frac{d\hat{H}_s}{ds}=[ \hat{\eta}_s,\hat{H}_s].
\label{eq:srgflow}
\end{equation}
Although other generators have been used \cite{li11,dicaire14a}, a common choice for this operator is the commutator of the evolved Hamiltonian with the kinetic energy, $\hat{\eta}_s=[\hat{T},\hat{H}_s]$.  %Here $[\hat{T},\hat{H}_s]$ is known as the generator for the transformation and $\hat{T}$ is the kinetic energy operator. 
This drives the Hamiltonian towards diagonal form in momentum space, thus decoupling high- and low-momentum states. The spread of the residual off-diagonal strength can be measured by the parameter with units of momentum $\lambda$ [where $s^{-1}=(\hbar\lambda)^4/M_N^2$], which can be used to follow the evolution of the Hamiltonian in place of $s$.  As $\lambda$ decreases, the Hamiltonian undergoes more evolution while $\lambda=\infty$ corresponds to the initial Hamiltonian.
%Note that $\lambda$ starts at infinity and approaches zero as the flow equation is evaluated.
%As the transformation is performed the momentum space Hamiltonian is driven to a band diagonal form \cite{glazek93}, thus decoupling the  high- and low-momentum sectors. The parameter $\lambda$ represents the momentum scale at which this decoupling occurs. 

Working within a discrete basis, %such as the Jacobi-coordinate HO wavefunctions adopted in this work, 
%the operator equation
Eq.~(\ref{eq:srgflow}) can be cast into a set of coupled first-order differential equations for the matrix elements of the flowing Hamiltonian $\hat H_s$, %and the right-hand side of the equation can be evaluated by simple matrix multiplications
with the right-hand side of the equation being simply given by matrix multiplications.  The procedure to determine the two- and three-body components of the evolved Hamiltonian within %this type of basis 
the Jacobi-coordinate HO wavefunctions adopted in this work was presented in Refs.~\cite{jurgenson09a,jurgenson11}.  In particular, depending on the absence or presence of $V^{3N}$ in %the initial Hamiltonian of
Eq.~(\ref{eq:ham}), one can identify three classes of evolved Hamiltonians: (1) $NN$-only, two-body Hamiltonian from the SRG evolution of the $NN$ force in the two-nucleon space; (2) $NN+3N$-induced, three-body Hamiltonian from the SRG evolution of the $NN$ force in the three-nucleon space; and (3) $NN+3N$, SRG Hamiltonian obtained from evolving the $NN$ plus initial $3N$ force in the three-nucleon system.

The consistent application of the SRG approach requires that any other operator, $\hat O$, %$\hat O^{(k\tau)}$, of rank $k$ in angular momentum and $\tau$ in isospin, 
undergo the same unitary transformation as the Hamiltonian, i.e. 
\begin{equation}
\hat{O}_s=\hat{U}_s\hat{O}_{s=0}\hat{U}_s^{\dagger}\,.
\label{eq:srgO}
\end{equation}
While this can be rewritten into a similar form as Eq.~(\ref{eq:srgflow}), it is more computationally efficient to compute the unitary transformation, $\hat{U}_s$, using the eigenvectors of the Hamiltonian before and after the transformation, $\ket{\psi_\alpha(0)}$ and $\ket{\psi_\alpha(s)}$ respectively,
\begin{equation}
\hat{U}_s=\sum\limits_{\alpha}\ket{\psi_\alpha(s)}\bra{\psi_\alpha(0)}.
\label{eqn:unitarytransform}
\end{equation}
In a discrete basis, the transformation of Eq.~(\ref{eq:srgO})  is then given, once again, by simple matrix multiplications. In particular,  for parity-conserving rank-zero operators (as for the Hamiltonian, working in the isospin formalism) $\hat U_s$ corresponds to a block-diagonal matrix with respect to the various angular-momentum, parity and isospin channels $(J^\pi T)$ of the system,  and the evolution can be performed block by block in parallel to that of $\hat H_s$. This type of evolution for operators, in both the $A=2$ and $A=3$ systems, has been recently implemented working within the Jabobi-coordinate NCSM basis~\cite{schuster14a}. The situation is more complicated for nonscalar operator, as they will couple different blocks. In this case, the unitary transformation must to be computed and stored for each block during the evolution of the Hamiltonian and the matrix elements of the evolved operator must be reconstructed in a second step. In this work, we have implemented this process in the $A=2$ space, while we defer to future work the technically more challenging process of evolving nonscalar operators in the three-body space.
 
%For a general operator, the unitary transformation must be computed for both the initial  and the final state of the operator, $\hat{U}_s^i$ and $\hat{U}_s^f$ respectively, giving a general matrix equation equation for the transformation,
%\begin{equation}
%\hat{O}_s^{JT}=\hat{U}_s^{i}\hat{O}_{s=0}^{JT}\hat{U}_s^{f\dagger}.
%\label{eqn:nonscalarevolve}
%\end{equation}
%For a scalar operator $\hat{U}_s^{i}=\hat{U}_s^{f}=\hat{U}_s$, so this equation can be simplified to a more familiar form,
%\begin{equation}
%\hat{O}_s=\hat{U}_s\hat{O}_{s=0}\hat{U}_s^{\dagger},
%\end{equation}

In general, to determine the two- and three-body components of an evolved operator we follow a similar procedure as that adopted for the Hamiltonian in Refs.~\cite{jurgenson09a,jurgenson11}. %we can say this later: Because we work in relative coordinates, all operators considered here are written as two-body operators. 
We start by evolving $\hat{H}_s$, hence calculating $\hat{U}_s$, in the $A=2$ system and determining the matrix elements of the two-body evolved operator, $\langle\hat{O}^{(2)}_s\rangle$,
%$= \langle\hat O_s\rangle$
through Eq.~(\ref{eq:srgO}). Next, (for scalar operators) we repeat the operation in the $A=3$ system, thus computing $\langle\hat{O}^{(3)}_s\rangle$, and then isolate the induced three-body %matrix elements 
components of the evolved operator via subtraction, $\langle\hat{O}^{(3)}_s\rangle-\langle\hat{O}^{(2)}_s\rangle$, where the second term corresponds to the two-body evolved operator embedded in the three-nucleon basis. This allows us to accurately calculate and separate the two- and three-body matrix elements 
of the evolved operator, which we can then use unchanged in calculations for any nucleus. %$^4$He. 
The second step can also be performed with or without the initial three-nucleon force in the Hamiltonian. Similar (but not quite parallel) to the three classes of Hamiltonian discussed earlier, this procedure leads to the following three stages of operator evolution: (1) Bare or unevolved operator; (2) 2B evolved, SRG-evolution of the operator in the two-body space; and (3) 3B evolved, SRG-evolution of the operator in the three-body space, allowing the induction of three-body terms.

%\subsection{Dipole Response and Electric Dipole Polarizability}
\subsection{Photoabsorption cross section and electric polarizability}
\label{sec:obs}
 At low excitation energies, when the long wavelength limit applies, %of the incident radiation is much larger than the spatial extension of the system under consideration, 
 the nuclear photoabsorption process can be described by the cross section~\cite{levinger50}
\begin{equation}
\label{eq:cross}
\sigma_\gamma(\omega)=4\pi^2\frac{e^2}{\hbar c}\omega R(\omega),
\end{equation}
where $\omega$ is the perturbing photon energy and $R(\omega)$ is the inclusive response function, given by,
\begin{equation}
\label{eq:response}
R(\omega)=\int d\Psi_f\left|\bra{\Psi_f}\hat{D}\ket{\Psi_0}\right|^2\delta(E_f-E_0-\omega),
\end{equation}
where $E_f$ and $E_0$  represent the final-state and g.s.\ energies along with their associated wavefunctions, $\ket{\Psi_f}$ and $\ket{\Psi_0}$, respectively and $\hat{D}$ is the electric dipole operator,
\begin{equation}
\label{eq:dipole}
\hat{D}=\sqrt{\frac{4\pi}{3}}\sum^A_{i=1}\frac{\tau^z_i}{2}r_iY_{10}(\hat{r}_i)\,.
\end{equation}
Here, $\tau^z_i$ is the third component of isospin and $\vec{r}_i = r_i \hat r_i$
is the position vector of the $i$th particle in the center-of-mass frame.
%To obtain the total cross section we: (i) solve the many-body Schr\"{o}dinger equation for the ground state, $\ket{\Psi_0}$, of $^4$He, and obtain the inclusive response, Eq. (\ref{eq:response}) by (ii) evaluation \cite{marchisio03} and (iii) inversion \cite{efros99a} of its integral transform with a Lorentzian kernel, described in more detail below, and (iv) calculate the photoabsorption cross section using Eq. (\ref{eq:cross}). 
%To compute the inclusive cross section we need the inclusive response function given in Eq. \ref{eq:response}.
 
To bypass the direct calculation of the final states, which for a light nucleus such as $^4$He are all in the energy continuum, the LIT method~\cite{efros94a,efros07a} %\cite{marchisio03} 
obtains the response function, $R(\omega)$, after the evaluation and subsequent inversion~\cite{efros99a,andreasi05a}  of its convolution with a Lorentzian kernel of finite width $\sigma_I$,
\begin{equation}
\label{eq:Lkernel}
L(\sigma_R,\sigma_I)=\int d\omega\frac{R(\omega)}{(\omega-\sigma_R)^2+\sigma_I^2}\,,
\end{equation}
where $\sigma_R$ is a continuous variable with unit of energy.
Taking advantage of the completeness of the eigenstates of the Hamiltonian this can be rewritten %in terms of a Green function with complex energy $z=E_0+\sigma_R+\sigma_I$ evaluated over $\hat D|\Psi_0\rangle$
as~\cite{marchisio03}
%We use the Lanczos algorithm \cite{haydock74, marchisio03} to write the LIT as a continued fraction,
\begin{equation}
L(\sigma_R,\sigma_I)=%-\frac{\bra{\Psi_0}\hat{D}^\dagger\hat{D}\ket{\Psi_0}}{\sigma_I}\text{Im} \{G(z)\},
-\frac{M_0}{\sigma_I}\text{Im} \{G(z)\},
\label{eq:Lgf}
\end{equation} 
where  $G(z)$ is the Green's function of Eq.~(\ref{eq:gf}) evaluated at the %$a_n$ and $b_n$ are the Lanczos coefficients and 
complex energy $z=E_0+\sigma_R+i\sigma_I$ on the starting Lanczos vector %$|\phi_0\rangle=\bra{\Psi_0}\hat{D}^\dagger\hat{D}\ket{\Psi_0}^{-1/2} \hat{D}\ket{\Psi_0}$.
$|\phi_0\rangle=M_0^{-1/2} \hat{D}\ket{\Psi_0}$. The quantity $M_0$ %$\bra{\Psi_0}\hat{D}^\dagger\hat{D}\ket{\Psi_0}$ 
is the total strength of the transition induced by the dipole operator, %The response function, $R(\omega)$, can then be obtained by the inversion of the integral transform of Eq. \ref{eq:Lkernel}. %For more information about the inversion process see references \cite{efros99}.
which can be either evaluated directly as the expectation value $M_0=\bra{\Psi_0}\hat{D}^\dagger\hat{D}\ket{\Psi_0}$ of the operator $\hat{D}^\dagger\hat{D}$ on the g.s.\ wavefunction, or as the  square norm $M_0=||\hat{D}\ket{\Psi_0}||^2$ of the vector $\hat D\ket{\Psi_0}$. In the first case, only the scalar component of the  $\hat{D}^\dagger\hat{D}$ operator is needed for the evaluation of the total dipole strength on the $J^\pi T=0^+0$ g.s. of the $^4$He nucleus. 

Similarly, in the unretarded dipole long-wavelength approximation adopted here,  the electric dipole polarizability of the nucleus is given by 
\begin{equation}
\alpha_E=2\frac{e^2}{\hbar c} \int d\Psi_f \frac{\left|\bra{\Psi_f}\hat{D}\ket{\Psi_0}\right|^2}{E_f-E_0},
%\sum_{n\neq0}\frac{|\bra{n}\hat{D}\ket{0}|}{E_n-E_0},
\end{equation}
which %where $\alpha$ is the fine-structure constant, $\ket{0}$ and $\ket{n}$ are the ground state and  the $n$th excited state, respectively, with their associated energy, $E_0$ and $E_n$, and $\hat{D}$ is the dipole operator. This can be rewritten as a sum rule,
corresponds to the double inverse-energy weighted sum rule of the photoabsorption cross action of Eq.~(\ref{eq:cross})
%\begin{equation}
%\alpha_E\equiv\frac{\sigma_{n}}{2\pi^2},
%\end{equation}
%with $n=-2$ for the polarizability. $\sigma_{-2}$ is then given by, 
\begin{equation}
\alpha_E=\frac{1}{2\pi^2}\int_{\omega_\text{th}}^{\infty}d\omega\frac{\sigma_\gamma(\omega)}{\omega^2},
\label{eq:integral}
\end{equation}
with %$\sigma_\gamma^{\text{ud}}(\omega)$ is the photoabsorption cross section of the unretarded dipole photons with energy $\omega$ and 
$\omega_{\text{th}}$ the threshold energy for photoabsorption. While the electric polarizability can be obtained through Eq.~(\ref{eq:integral}) by numerical integration of the computed cross section of Eq.~(\ref{eq:cross}), it is more efficient and numerically more accurate to take advantage of the completeness of the eigenstates of the Hamiltonian and directly evaluate it by means of the Lanczos method as   
\begin{equation}
\alpha_E=-2\frac{e^2}{\hbar c} %\bra{\Psi_0}\hat{D}^\dagger\hat{D}\ket{\Psi_0}G(E_0),
M_0 G(E_0)
\label{eq:agf}
\end{equation} 
with the same starting vector as in Eq.~(\ref{eq:Lgf}).

%As a final observation, we note that the total dipole strength $\bra{\Psi_0}\hat{D}^\dagger\hat{D}\ket{\Psi_0}$ can be either evaluated directly as the expectation value of the operator $\hat{D}^\dagger\hat{D}$ on the g.s. wavefunction, or as $||\hat{D}\ket{\Psi_0}||^2$. In the first case, only the scalar component of the transition operator is needed for the evaluation of the total dipole strength on the $J^\pi T=0^+0$ g.s. of the $^4$He nucleus. 

\section{Results}
%Although the focus of this work are the photoabsorption cross section and electric dipole polarizability of the $^4$He nucleus, in the following we will also examine other quantities calculated in the two- and three-nucleon systems. 
All results are obtained employing  the Idaho %$500$MeV 
N$^3$LO nucleon-nucleon interaction of Ref.~\cite{entem03a} and the N$^2$LO three-nucleon force from Ref.~\cite{navratil07a} with the low energy constants adjusted to reproduce the triton half-life and the binding energies of $^3$H and $^3$He nuclei~\cite{gazit09a}. Unless otherwise stated, we truncate all of our calculations in the $A=2$ model space at $N_\text{max}=300$ and the $A=3$ model space at $N_\text{max}=40$, denoted as $N_\text{A2max}$ and $N_\text{A3max}$, respectively. The HO model space size for the $^4$He system will be simply indicated as $N_{\rm max}$.

In Sec.~\ref{sec:Dev} we start by exploring the evolution of a few matrix elements of the dipole transition. Next, in Sec.~\ref{sec:conv}, we discuss %Because of the large model spaces accessible with the Jacobi-coordinate basis,  we can obtain accurately 
the convergence properties of our results with respect to variations in both $N_{\rm max}$ %the that are robust against variation of the 
and HO frequency, $\hbar\Omega$. Finally, in Sec.~\ref{sec:lam}, we study the $\lambda$ dependence of our calculations and,  in Sec.~\ref{sec:expt} present a comparison with available experimental data.

\subsection{Two-body evolved dipole operator}
\label{sec:Dev}
\begin{figure*}
\begin{center}
\includegraphics[width=0.9\textwidth]{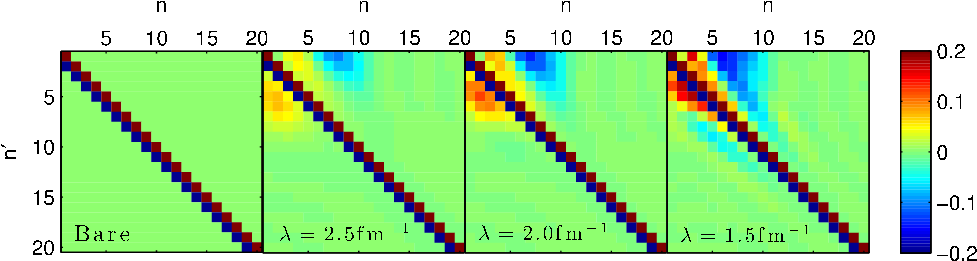} 
\end{center}
\caption{(Color online) SRG evolution of the two-body dipole operator in HO space for the  $^3$S$_1$ to $^3$P$_2$ transition. The color bar represents the value of the dipole matrix elements and is truncated to highlight the off-diagonal behavior as a function of evolution, from bare ($\lambda=\infty$) to $\lambda=1.5$ fm$^{-1}$. The matrix elements have units of fm.\label{fig:Devolve}}
\end{figure*}

%Our previous work on operator evolution had focused solely on scalar operators%in the two- \cite{anderson10} and three-body space 
%~\cite{schuster14a}. 
To obtain the photoabsorption cross section and electric dipole polarizability of Sec.~\ref{sec:obs} within the SRG approach, we need to consider the %two-body 
evolution of the electric dipole operator of Eq.~(\ref{eq:dipole}) %extend these investigations to the dipole operator, given by
%\begin{equation}
%\label{eq:dipole}
%\hat{D}=\sqrt{\frac{4\pi}{3}}\sum^A_{i=1}\frac{\tau^z_i}{2}r_iY_{10}(\vec{r}_i),
%\end{equation}
that induces a $J^{\pi}T=1^{-}1$ transition between initial and final states. 
%Indeed, while 
For $^4$He, the total dipole strength entering Eqs.~(\ref{eq:Lgf}) and (\ref{eq:agf}) can be evaluated as the expectation value of a scalar operator, and we can use the technology we developed in Ref.~\cite{schuster14a} to renormalize $\hat{D}^\dagger\hat{D}$ (a scalar operator) up to the three-body level. However, the matrix elements of $\hat D$ are still needed to compute the Lanczos starting vector, which is proportional to $\hat D\ket{\Psi_0}$. %In the present work, we will limit ourselves to evolving $\hat D$ in the $A=2$ space and defer the much more  
As already mentioned in Sec.~\ref{sec:srg}, properly evolving a nonscalar operator introduces additional technical complications, particularly in the $A=3$ system. At the same time, we expect that the renormalization of the dipole will have only a minor effect on the Green's functions of Eq.~(\ref{eq:Lgf}) and (\ref{eq:agf}) if the Hamiltonian is evolved up to the three-body level. Therefore, for the time being we will limit ourselves to two-body matrix elements of the evolved $\hat D$ in the calculation of the Lanczos starting vector.

Fig. \ref{fig:Devolve} shows snapshots of the evolution of the dipole operator in HO space for $^3$S$_1$ (T=0) to $^3$P$_2$ (T=1) transitions. The color bar represents the value of the HO matrix elements and is truncated to highlight the off diagonal behavior as the operator is evolved. Since this is a transition between different initial and final states, the representation in HO space is not symmetric. Snapshots of this kind are useful for examining the behavior of the matrix elements during evolution and have been shown previously for operators evolved in momentum space \cite{anderson10} and for the Hamiltonian evolved in HO \cite{Dicaire2014,roth14a} and momentum space \cite{bogner07,anderson08a}. Here, the discretized axes, $n$ and $n^\prime$, are the radial quantum numbers of the HO wavefuntion and directly correspond to the energy is HO space.
%As the operator is evolved, the strength in the off-diagonal matrix elements tends to increase. 
For this transition, the bare operator starts as a lower bidiagonal matrix and as $\lambda$ decreases we see increased strength in the off diagonal matrix elements.
%So rather than evolving the operator to a more diagonal form, as we seen with the Hamiltonain, the SRG evolution tends to spread out the operator in HO space.
So while the SRG evolves the momentum space Hamiltonian to a more diagonal form, it spreads out the dipole operator in HO space.

\subsection{Convergence}
\label{sec:conv}
\begin{figure}[b]
\includegraphics[width=0.45\textwidth]{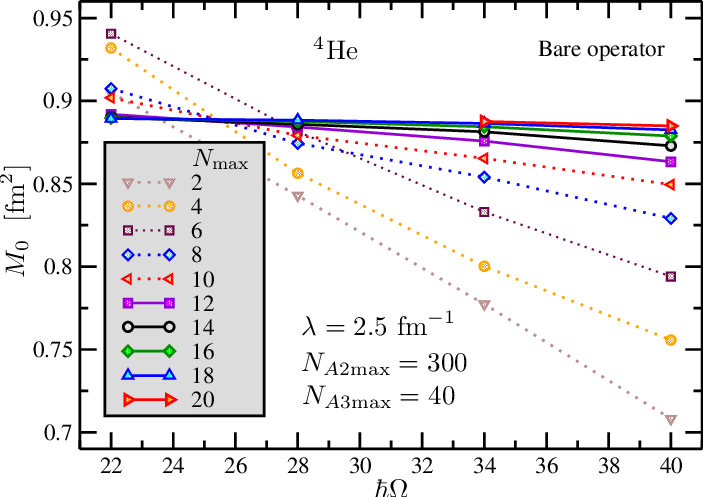} 
\caption{(Color online) Convergence of the total dipole strength $M_0$ of $^4$He as a function of $N_\text{max}$ using the bare operator and evolved wavefunctions from the $NN$+$3N$ Hamiltonian with $\lambda=2.5$ fm$^{-1}$ at $\hbar\Omega = 22$, $28$, $34$, and $40$ MeV.  \label{fig:oddo_converge_NNNhW}}
\end{figure}
\begin{figure*}[t]
\begin{minipage}[c]{.48\linewidth}
\centering
\includegraphics*[width=.94\linewidth]{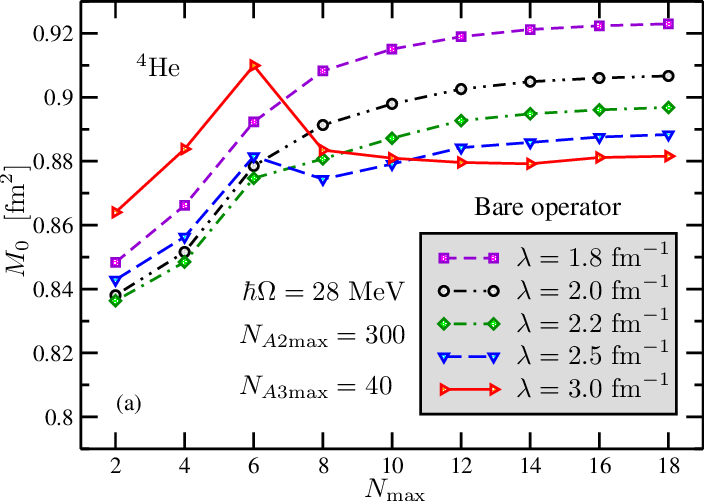}
\end{minipage}
\hfill
\begin{minipage}[c]{.48\linewidth}
\centering
\includegraphics*[width=.94\linewidth]{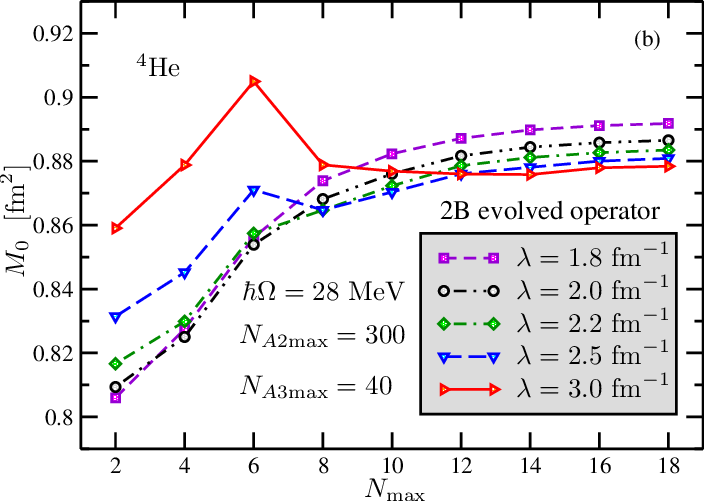}
\end{minipage}
\caption{
Convergence of the total dipole strength $M_0$ of $^4$He as a function of $N_\text{max}$ at $\hbar\Omega=28$ MeV using (a) the bare and (b) the 2B evolved $\hat D$ operator and wavefunctions from the $NN$+$3N$ Hamiltonian with $\lambda=1.8$, $2.2$, $2.5$, and $3.0$ fm$^{-1}$.  \label{fig:oddo_converge_NNNbare}}
\end{figure*}
In this section, we discuss the behavior of our calculations with respect to variations of the  frequency $\hbar\Omega$ and size $N_{\rm max}$ of the adopted HO model space. 

We start in Fig.~\ref{fig:oddo_converge_NNNhW} by analyzing the total strength, $M_0$, of the bare dipole operator evaluated on the $^4$He evolved g.s.\ wavefunction (using, in this example, the %SRG-evolved 
$NN$+$3N$ Hamiltonian with $\lambda = 2.5$ fm$^{-1}$) for a range of HO frequencies and various basis sizes. As $N_{\rm max}$ increases, the total dipole strength becomes more and more independent from the choice of the $\hbar\Omega$ value in the range $22-40$ MeV, reaching a flat behavior in the largest model spaces. The weakest $N_{\rm max}$ dependence is found for frequencies between 22  and 28 MeV, for which an excellent convergence is already achieved at $N_{\rm max}=18$ proceeding from above and from below, respectively. These two $\hbar\Omega$ values will be adopted for the reminder of our study. In addition, our choices for $N_\text{max}$ have been shown to be fully converged and robust against changes to the HO frequency \cite{CK13}.

The typical convergence of $M_0$ as a function of $N_\text{max}$,
computed as the norm $||\hat D|\Psi_0\rangle ||^2$, for the bare and two-body evolved dipole operators is presented in Figs.~\ref{fig:oddo_converge_NNNbare}(a) and \ref{fig:oddo_converge_NNNbare}(b), respectively. %acting on the evolved g.s.\ wavefunctions (using, in this example, the $NN$+$3N$ Hamiltonian) 
%for a range of $\lambda$ from $1.8$ to $3.0$ fm$^{-1}$ %is presented in Figs.~\ref{fig:oddo_converge_NNNbare}(a) and \ref{fig:oddo_converge_NNNbare}(b), respectively. %Panel (a) shows results obtained using the bare dipole operator, while the two-body evolved operator was utilized in Fig.~\ref{fig:oddo_converge_NNNbare}(b). 
As the dipole is a long range operator, we see almost no increase in the rate of convergence of the evolved over the bare operator (both evaluated, as in Fig.~\ref{fig:oddo_converge_NNNhW}, on $NN$+$3N$ evolved wavefunctions). Rather, the SRG evolution of the wavefunction provides a smooth convergence pattern, especially at smaller values of $\lambda$, regardless of the level of operator evolution. As an example, for $\lambda=2.5\text{ fm}^{-1}$ the $M_0$ values begin to follow an exponential convergence above $N_\text{max}=10$, whereas at $\lambda=1.8\text{ fm}^{-1}$ the exponential convergence already starts at $N_\text{max}\sim$ 6. This could be used effectively to extrapolate to $N_\text{max}=\infty$ in heavier systems where one cannot feasibly reach large $N_\text{max}$ values or where convergence of observables is very slow. 

As will be discussed in the next section and can be seen in Figs.~\ref{fig:oddo_converge_NNNbare}(a) and \ref{fig:oddo_converge_NNNbare}(b), for dipole transitions the converged values tend to increase as $\lambda$ decreases. This is due to the omission of induced many-body [three- and four-body in the case of Fig.~\ref{fig:oddo_converge_NNNbare}(b)] contributions to the SRG evolved operator. %Indeed, the $\lambda$ spread yielded by the 2B evolved operator is a factor of three smaller than for the bare operator.
Indeed, the difference between the $M_0$ values obtained with bare and 2B evolved operators is much larger at $1.8$ than at $3.0$ fm$^{-1}$ due to the increasing strength of the SRG induced terms as $\lambda$ decreases.
\begin{figure}[b]
\includegraphics[width=0.45\textwidth]{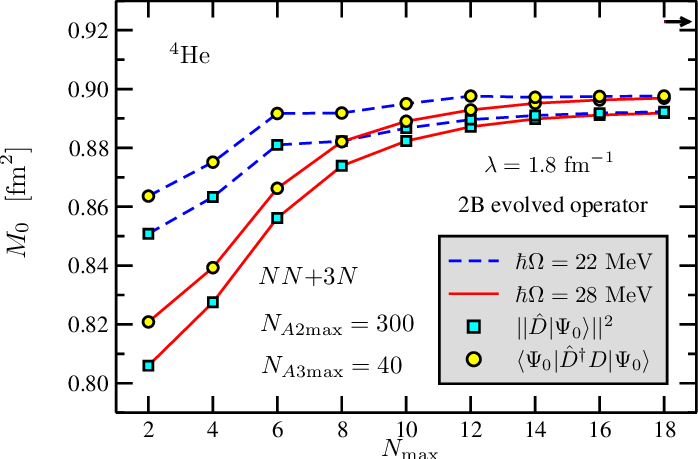} %bounding box issue!
\caption{(Color online) Convergence as a function of $N_\text{max}$ of the two-body evolved total dipole strength, $M_0$, of $^4$He computed as $||\hat{D}\ket{\Psi_0}||^2$ (squares) and $\hat{D}^\dagger \hat{D}$ (circles) for $\lambda=1.8 \text{ fm}^{-1}$ and $\hbar\Omega=22$ MeV (dashed lines) and $28$ MeV (solid lines). The arrow shows the converged value of $M_0$ computed with the bare operator.  Results were obtained using the wavefunction from the $NN$+$3N$ Hamiltonian. \label{fig:oddo_converge_NNN}}
%\caption{(Color online) Convergence of the bare (circles) and two-body SRG evolved (squares) total dipole strength of $^4$He as a function of $N_\text{max}$ for $\lambda=1.8 \text{ fm}^{-1}$, (a), and $2.5 \text{ fm}^{-1}$  (b). Results were obtainted using the wavefunction from the $NN$+$3N$ Hamiltonian with $\hbar\Omega=22$ MeV (dashed line) and $28$ MeV (solid line). \label{fig:oddo_converge_NNN}}
\end{figure}

%To conclude the discussion on the total dipole strength, 
In Fig.~\ref{fig:oddo_converge_NNN} we compare the convergence with respect to $N_\text{max}$ of $M_0$ computed in two different ways: as the norm $||\hat D|\Psi_0\rangle ||^2$ of the 2B evolved dipole operator, $\hat{D}$, acting on the $^4$He g.s.\ and as the expectation value on the g.s.\ wavefunction of the 2B evolved $\hat D^\dagger\hat D$ operator.   The two procedures yield the same result when the bare operators are employed, represented by the arrow in the figure. However, in general the same is not true upon the SRG evolution, which results in a different renormalization for operators exhibiting different short-range properties (in this case, $r$ versus $r^2$, respectively). Similar to what we have observed for the bare operator, varying the oscillator frequency from $22$ to $28$ MeV produces little change in the converged value of the observables. This is not surprising considering the large model spaces reached in the present work. More interesting are the differences in the size of 2B induced contributions for the total dipole strength calculated as $||\hat{D}\ket{\Psi_0}||^2$ versus $\bra{\Psi_0}\hat D^\dagger\hat D\ket{\Psi_0}$. A somewhat larger renormalization is observed in the case of the former, shorter-range operator.  
\begin{figure}[t]
\includegraphics[width=0.45\textwidth]{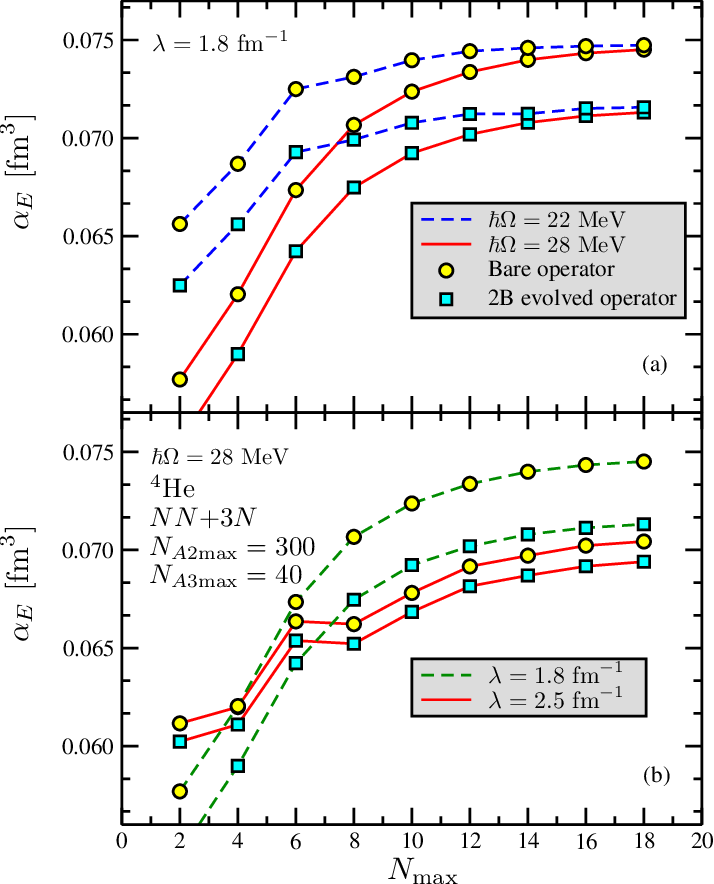} %bounding box issue!
\caption{(Color online) Convergence of the bare (circles) and two-body SRG evolved (squares) electric polarizability of $^4$He as a function of $N_\text{max}$ for (a), $\lambda=1.8 \text{ fm}^{-1}$ with $\hbar\Omega=22$ MeV (dashed line) and $28$ MeV (solid line), and (b), with fixed $\hbar\Omega=28$ MeV at $\lambda=1.8$ (dashed line) and $2.5 \text{ fm}^{-1}$ (solid line). Results were obtainted using the wavefunction from the $NN$+$3N$ Hamiltonian.   \label{fig:polarize_converge}}
\end{figure}

Next, in Fig.~\ref{fig:polarize_converge}, we consider the electric dipole polarizability, calculated according to Eq.~(\ref{eq:agf}) with $M_0=||\hat{D}\ket{\Psi_0}||^2$. Two values of the frequency ($\hbar\Omega=22$ and $28$ MeV) and SRG momentum scale ($\lambda = 1.8$ and $2.5$ fm$^{-1}$) are explored for $N_{\rm max}$ values varying between 2 and 18.  The convergence patterns obtained for the bare versus 2B evolved operator are once again very similar, although a slightly faster flattening of the curves can be observed for the latter, and the two frequencies adopted yield  very similar results at $N_{\rm max}=18$. As with the total dipole strength, the inclusion of the 2B evolved operator reduces the spread in the SRG momentum scale and the contribution of the two-body induced terms is larger for $\lambda=1.8$ fm$^{-1}$.

\begin{figure}[t]
\includegraphics[width=0.45\textwidth,clip=true]{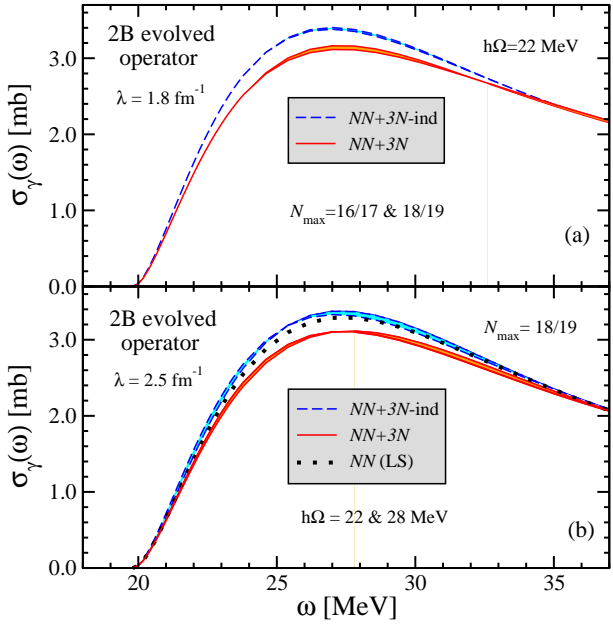} %bounding box issue!
\caption{(Color online) Dependence of the $^4$He total photoabsorption cross section computed with the $NN$+$3N$-induced (region delimited by dashed [blue] lines) and $NN$+$3N$ (region delimited by solid [red] lines) Hamiltonians and 2B evolved dipole operator on: (a) the model space size $N_{\rm max}$ at $\hbar\Omega = 28$ MeV and $\lambda=1.8$ fm$^{-1}$; and (b) the HO frequency $\hbar\Omega$ at $N_{\rm max}=18/19$ and $\lambda = 2.5$ fm$^{-1}$. Also shown (dotted [black] line) is the result of the LS calculation of Ref.~\cite{quaglioni07} using the N$^3$LO $NN$ interaction.   \label{fig:cs_converge}}
\end{figure}
To conclude this section, we assess by means of Fig.~\ref{fig:cs_converge} the sensitivity of the $^4$He photoabsorption cross section, computed according to Eq.~(\ref{eq:cross}), to variations of the HO model space size and frequency. The total dipole strength entering the evaluation of the LIT~(\ref{eq:Lgf}), and hence of the response function $R(\omega)$ of Eq.~(\ref{eq:response}), was obtained as $M_0=||\hat D|\Psi_0\rangle ||^2$ using the 2B evolved operator. Both $NN$+$3N$-induced and $NN$+$3N$ Hamiltonians are considered. For the sake of comparison, after being computed, all theoretical cross sections are shifted to the experimental threshold for the $^4$He photo-disintegration, $E_{\rm th}$ = 19.8 MeV ($\omega\rightarrow\omega+\Delta E_{\rm th}$, with $\Delta E_{\rm th}$ being the difference of the calculated and experimental thresholds).  This allows us to highlight differences beyond those occurring at the level of the $^4$He and $^3$H binding energies. Due to the selection rules associated with the dipole operator~(\ref{eq:dipole}), for a given $N_{\rm max}$ in the $J^\pi T=0^+0$ model space used to expand $\ket{\Psi_0}$, a complete calculation of Eq.~(\ref{eq:Lgf}) requires the expansion of the starting Lanczos vector $\ket{\varphi_0}=M_0^{-1/2}\hat D\ket{\Psi_0}$ over a $J^\pi T = 1^{-}1$ space up to $N_{\rm max}+1$. This is the origin of the odd/even notation for $N_{\rm max}$ introduced in Fig.~\ref{fig:cs_converge}.  The relative uncertainty due to the finite size of the HO space, estimated from the difference of the cross section calculated at $N_{\rm max} = 18/19$ and $16/17$ is largest for the $NN$+$3N$ Hamiltonian, remaining below 2\% above $\omega\sim22$ MeV. At lower energies -- where the cross section is smaller -- the relative uncertainty grows somewhat reaching a value of $\sim 8\%$ at threshold. Varying the HO frequency from 28 to 22 MeV produces results within $3\%$, except for energies very close to threshold. Finally, as shown in Fig.~\ref{fig:cs_converge}(b), the present $NN$+$3N$-induced results are consistent with those obtained in Ref.~\cite{quaglioni07} using a LS transformation of the N$^3$LO $NN$ potential at the three-body cluster level, in which the dipole operator was not renormalized.   
\begin{figure}[t]
\begin{center}
\includegraphics[width=0.45\textwidth]{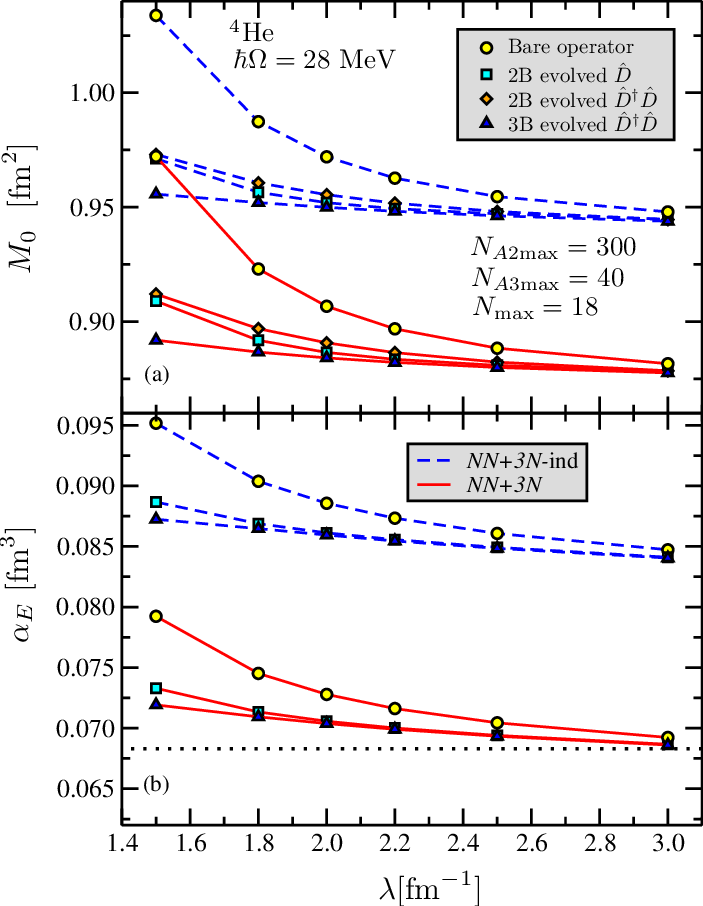} %bounding box issue!
\end{center}
\caption{(Color online) Dependence of (a) total strength of the dipole transition and (b) electric dipole polarizability on variations of the SRG flow parameter, $\lambda$, for $N_{\text{max}}=18$ and $\hbar\Omega=28$ MeV, obtained using wavefunctions from the $NN+3N$-induced (dashed lines) and $NN$+$3N$ (solid lines) Hamiltonians along with four types of operators:  bare (circles), 2B evolved $\hat D$ (squares), 2B evolved $\hat D^\dagger\hat D$ (diamonds) and 3B evolved $\hat D^\dagger\hat D$ (triangles). The dotted line in panel (b) indicates the evaluation of Ref.~\cite{stetcu09a} based on a LS renormalization of the N$^3$LO $NN$ plus N$^2$LO $3N$ interactions and bare dipole operator. See the text for more details. %Only the $M_0$ component of the polarizability was evolved up to the three-body level [see Eq.~(\ref{eq:agf}) and the text for more information].
\label{fig:lambdastudy}}
\end{figure}

\begin{figure*}[t]
\begin{center}
\includegraphics[width=0.7\textwidth, clip=true]{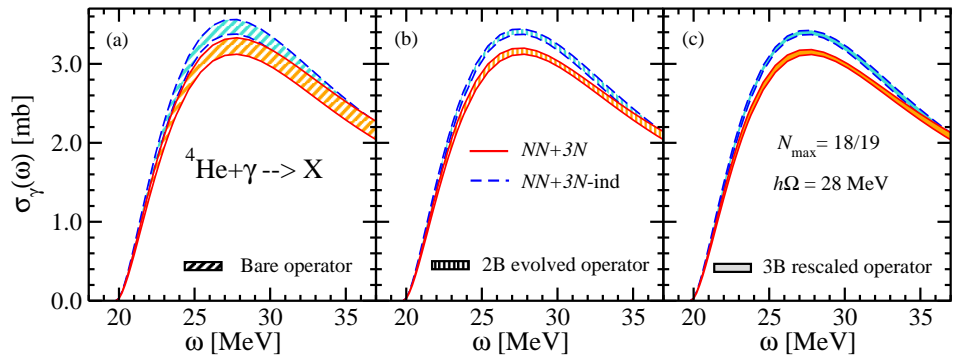} %bounding box issue!
\end{center}
\caption{(Color online) Dependence (represented as the width of the bands) on the variation of $\lambda$ between $1.8$ and $3.0$ fm$^{-1}$ of the $^4$He photo-absorption cross section, $\sigma_\gamma(\omega)$, as a function of the photon energy, $\omega$, at  $N_{\text{max}}=18/19$ and $\hbar\Omega=28$ MeV, using the $NN+3N$-induced (dashed contours) and $NN+3N$ (solid contours) wavefunctions.  Calculations were obtained with: (a) the bare dipole operator;  (b) the 2B evolved dipole operator; and (c) rescaling the 2B evolved results by the ratio $\bra{\Psi_0}\hat D^\dagger\hat D\ket{\Psi_0}/||\hat D|\Psi_0\rangle ||^2$, with the $\hat D^\dagger\hat D$ operator evolved in the three-nucleon space (3B rescaled operator, see text for details). \label{fig:cslambda}}
\end{figure*}
\subsection{SRG resolution scale dependence}
\label{sec:lam}

\begin{table*}[t]
\caption[]{Calculated $^4$He g.s.\ energy $E_0$, point-proton root-mean square radius $\sqrt{\langle r^2_p\rangle}$, total dipole strength $\bra{\Psi_0}\hat D^\dagger\hat D\ket{\Psi_0}$, and electric dipole polarizability $\alpha_E$ using the using the $\lambda=1.8$ and $3.0$ fm$^{-1}$ $NN+3N$-induced\ and $NN+3N$ Hamiltonians along with three-body evolved operators compared to results published in the literature and experiment. See the text for more details.
\label{table:results}}
\begin{ruledtabular}
\renewcommand{\arraystretch}{1.3}
\begin{tabular}{l  c  c c  c c c  c  }
Interaction& $\lambda$ (fm$^{-1}$)&$E_{\rm g.s.}$ (MeV) &  $\sqrt{\langle r^2_p\rangle}$ (fm) & $\bra{\Psi_0}\hat D^\dagger\hat D\ket{\Psi_0}$ (fm$^2$ ) & $\alpha_E$ (fm$^3$ ) \\
%Units& MeV& fm & fm$^2$ & fm$^3$\\
\hline
$NN$+$3N$-ind       &$1.8$   &  $-25.325(1)$                  &   $1.5231(11)$                    &$0.9520(3)$                   & $~0.08647(5)\phantom{(1)}$\\
                                 &3.0       &$-25.348(2)$                    &   $1.5165(12)$                     &$0.9439(4)$                  &  $~0.08404(5)\phantom{(1)}$\\
N$^3$LO $NN$ (LS)~\cite{quaglioni07}&--          & $-25.39(1)\phantom{1}$  &  $1.515(2)\phantom{11}$   &$0.943(1)\phantom{1}$ & -- \\
 $NN$+$3N$            &$1.8$   & $-28.464(2)$                  &   $1.4723(7)\phantom{1}$    &$0.8867(4)$                  & $~0.07093(5)\phantom{(1)}$\\
                                &3.0        & $-28.458(3)$                  &  $1.4651(5)\phantom{1}$     &$0.8776(5)$                  & $~0.06861(5)\phantom{(1)}$\\
% + N$^2$LO $3N$ (LS) &-- & $-28.50(2)\phantom{1}$ & $-$& $-$ & $-$\\
% AV18  &-- & -30.84 & $-$ & $-$ & $-$\\
% + UIX &-- & -30.84 & $-$ & $-$ & $-$\\
Evaluation (LS)~\cite{stetcu09a}        &--          & --                                    & --                					 &--                                  &$~0.0683(8)(14)$ \\
Expt.                        &--          & $\phantom{1}-28.296$ \cite{tilley92a}& \phantom{11}$1.455(7)$~\cite{radius}& --                               & $0.072(4)$~\cite{Friar1977}\\
                                &            &                                        &                     					&                                   & $0.076(8)$~\cite{Pachucki2007}\\
\end{tabular}
\end{ruledtabular}
\end{table*}
In Fig. \ref{fig:lambdastudy}, we study the dependence on the SRG evolution parameter of the $^4$He total dipole strength and electric dipole polarizability.  These results where obtained with an oscillator frequency of $\hbar\Omega=28$ MeV and converged calculations at $N_{\text{max}}=18$. %The range of $\lambda$ values that we present, $1.5\text{ fm}^{-1}$ to $3.0\text{ fm}^{-1}$, has been shown to provide a good balance between improved convergence and growth of the SRG induced terms in energy calculations~\cite{jurgenson09a, jurgenson11} of $A\le10$ nuclei.

The behavior of the total dipole strength as a function of $\lambda$, presented in Fig.~\ref{fig:lambdastudy}(a), is consistent with %what has been seen 
that obtained in our previous study~\cite{schuster14a} of the evolution of the $\hat D^\dagger\hat D$ operator up to the three-body level. Different from that work, here we also show results obtained by computing $M_0$ as the norm $||\hat D|\Psi_0\rangle ||^2$ of the two-body evolved dipole operator acting on the g.s.\ wavefunction. When using the bare operator, the observables have a significant dependence on $\lambda$, particularly at smaller values. When using the two-body evolved operators, this dependence is reduced. The difference between the bare and two-body evolved operator, which we refer to as the two-body contribution to the evolution, is larger at smaller values of $\lambda$ and tends to decrease rapidly as $\lambda$ increases. Further, such two-body contribution  is found to be larger when the total strength is calculated as $||\hat D|\Psi_0\rangle ||^2$ using the two-body evolved dipole operator. This is related to the longer range of the $\hat D^\dagger\hat D$ operator compared to the dipole itself.  For the time being, results for the evolution at the three-body level have been obtained only for the scalar $\hat D^\dagger\hat D$ operator~\cite{schuster14a}. 
The three-body contribution to the operator evolution is much smaller than the two-body contribution, establishing a hierarchy in the magnitude of the SRG induced terms for operator evolution. Overall, the smallest spread in $\lambda$ is found using the three-body evolved $\hat D^\dagger\hat D$ operator. The slight residual dependence on $\lambda$ is due to the induced four-body terms that we do not take into account for these calculations.  

The electric dipole polarizabilty, presented in Fig.~\ref{fig:lambdastudy}(b), shows a similar trend to that of the total dipole strength. The inclusion of the two-body induced terms of the operator provides a substantial correction to the polarizability, especially at smaller values of $\lambda$.  To estimate the contribution to this observable of three-body induced terms of the operator, in Fig.~\ref{fig:lambdastudy}(b) we also show the polarizability (triangles) obtained by rescaling the 2B evolved polarizability (squares), by the ratio $\bra{\Psi_0}\hat D^\dagger\hat D\ket{\Psi_0}/||\hat D|\Psi_0\rangle ||^2$, where the $\hat D^\dagger\hat D$ operator is evolved in the three-nucleon space and is $||\hat D|\Psi_0\rangle ||^2$ evolved in the two-nucleon space.   %For the time being we are limited  to matrix elements of $\hat D$ evolved at most in the two-nucleon system when computing the Lanczos starting vector entering the Green's function $G(E_0)$ of Eq.~(\ref{eq:agf}). Assuming that 
The residual dependence on $\lambda$ displayed by these rescaled results comes then from four-body induced SRG terms but also from missing three-body induced dipole operator terms in the calculation of the Green's function, $G(E_0)$, of Eq.~(\ref{eq:agf}).  This latter contribution is expected to be small if the Hamiltonian is evolved up to the three-body level. Also shown in the figure as a dotted line is the evaluation of Ref.~\cite{stetcu09a} based on a LS renormalization of the N$^3$LO NN plus N$^2$LO $3N$ interactions and bare dipole operator.

Finally, in Fig.~\ref{fig:cslambda} we explore the effect of the SRG evolution of the transition operator on the $^4$He photoabsorption cross section. %as a function of the photon energy, $\omega$. 
This study was performed using our largest model space of $N_\text{max}=18/19$ at $\hbar\Omega=28$ MeV and both $NN$+$3N$-induced\ and $NN$+$3N$ wavefunctions, varying the SRG resolution scale between $1.8$ and $3.0$ fm$^{-1}$. We choose this range of $\lambda$ because previous structure calculations show that the g.s. energy is mostly independent of the transformation in this region. As shown in Fig.~\ref{fig:cslambda}(a), when using the bare dipole operator there is a clear dependence of the cross section on $\lambda$, and the spread is slightly larger for the calculation using the $NN$+$3N$ Hamiltonian. Specifically, beginning at a photon energy of $26$ MeV and persisting up to the largest energy shown here there is a spread of more than $0.2$ mb between the $NN$+$3N$ cross sections obtained with the smallest and largest value of  $\lambda$  (corresponding respectively to the upper and lower bounds of the shaded areas). This amounts to an effect between 6 and 11\%, depending on the photon energy, which is substantially larger than our uncertainty due to the finite size of the HO model space or choice of frequency. Further, this spread is comparable to the contribution coming from the inclusion of the initial chiral $3N$ force into the Hamiltonian,  which -- at a given $\lambda$ value -- quenches the peak of the cross section by about $0.25$ mb.  When we evolve the dipole operator in the two-body space [see Fig.~\ref{fig:cslambda}(b)], the spread in the cross section is a factor of three tighter, about $0.06$ mb (between 2\% and 4\% in the range 24 MeV $\le\omega\le35$ MeV), and the effect of the inclusion of the initial chiral $3N$ force can be clearly singled out. To take into account three-body induced terms of the transition operator, at least in part, the cross sections of Fig.~\ref{fig:cslambda}(b) can be further rescaled by the ratio $\bra{\Psi_0}\hat D^\dagger\hat D\ket{\Psi_0}/||\hat D|\Psi_0\rangle ||^2$, with the $\hat D^\dagger\hat D$ operator evolved in the three-nucleon space (3B rescaled operator). The result of this operation, shown in Fig.~\ref{fig:cslambda}(c), is mainly an overall small reduction of all curves, and a very minor narrowing of the spread in $\lambda$. The remaining $\lambda$ dependence is due, once again, to four-body induced SRG terms and from missing three-body induced dipole operator terms in the calculation of the Green's function, $G(E_0)$, of Eq.~(\ref{eq:Lgf}).

\subsection{Comparison with literature and experiment}
\label{sec:expt}
Table~\ref{table:results} presents a summary of our results for total dipole strength $\bra{\Psi_0}\hat D^\dagger\hat D\ket{\Psi_0}$ and electric dipole polarizability $\alpha_E$ obtained employing the $NN+3N$-induced and $NN+3N$ Hamiltonians along with the three-body evolved $\hat D^\dagger\hat D$ operator in the largest model space. For the electric polarizability, these results represent an upper bound as the effect of three-body induced dipole operator terms in the calculation of the Green's function of Eq.~(\ref{eq:agf}) are still missing. Two values of $\lambda$,  $1.8$ and $3.0$ fm$^{-1}$, are shown to help quantify the effect of missing induced terms. For completeness, we also show the corresponding values of the g.s.\ energy, $E_0$, and point-proton root-mean square radius, $\sqrt{\langle r^2_p\rangle}$, of Ref.~\cite{schuster14a}, including three-body induced terms. The errors estimates of the observables are computed as the difference between the value at largest model space, $N_\text{max}=18$, and the next smallest model space, $N_\text{max}=16$. The present results for the g.s.\ energy are the same as the previous NCSM calculation of Ref.~\cite{jurgenson11} and those for the $NN$+$3N$-induced point-proton radius and total dipole strength are consistent with those obtained in Ref.~\cite{quaglioni07} using a LS transformation of the N$^3$LO $NN$ potential at the three-body cluster level, in which the operators were not renormalized. In particular, the agreement with the LS values is excellent for $\lambda=3.0$ fm$^{-1}$, where the contribution of four-body induced terms is negligible. A similar comparison for the $NN$+$3N$ Hamiltonian is not possible, because the results of Ref.~\cite{quaglioni07} were obtained with a sightly different parameterization of the N$^2$LO three-nucleon force. Also in very good agreement with the evaluation of Ref.~\cite{stetcu09a} and with experiment is the electric dipole polarizability computed with the $NN$+$3N$ interaction.

\begin{figure}[t]
\begin{center}
\includegraphics[width=0.45\textwidth,trim=0cm 0cm 0cm 0cm,clip=true]{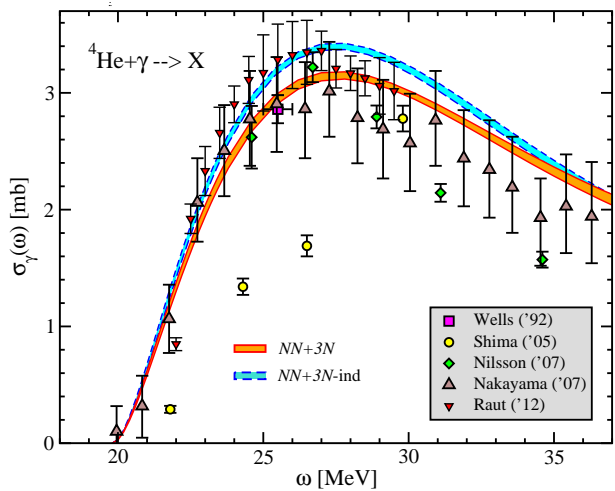}
\end{center}
\caption{(Colar online) The $^4$He photo-absorption cross section as a function of excitation energy, $\omega$, for $NN+3N$-induced (blue dashed line) and $NN$+$3N$ (red solid line) interactions with a model-space truncation of $N_{\text{max}}=18/19$ and oscillator frequence of $\hbar\Omega=28$ MeV. The total cross section is compared to experimental results from: Wells \cite{wells92}, Shima \cite{shima05}, Nilsson \cite{nilsson07}, Nakayama \cite{nakayama07} and Raut \cite{Raut2012}. See the text for more details. \label{fig:csexpt}}
\end{figure}

For completeness, in Fig.~\ref{fig:csexpt}, we compare our results for the $^4$He  photoabsorption cross section of Fig.~\ref{fig:cslambda}(c) to experimental data in the region $\omega < 40$ MeV, where corrections to the unretarded dipole approximation used here to describe the photo disintegration process are expected to be largely negligible. As for the electric polarizability, the present results represent an upper bound due to the missing effect of three-body induced dipole operator terms in the calculation of the Green's function of Eq.~(\ref{eq:agf}).   The photodisintegration of $^4$He has been the subject of many experiments (see, e.g. Refs.~\cite{shima05}, \cite{nilsson07}, \cite{nakayama07}, and \cite{Raut2012} for the most recent ones) and has already been investigated in {\em ab initio} calculations including three-nucleon forces~\cite{gazit06,quaglioni07}. The results obtained  here with the $NN$+$3N$-induced Hamiltonian are close the recent Coupled Cluster calculation of Ref.~\cite{bacca14}, using the bare N$^3$LO potential.  Different from Ref.~\cite{quaglioni07}, here the $NN$+$3N$ results have been obtained starting from the N$^2$LO $3N$ force of Ref.~\cite{gazit09a}. Therefore, the two calculation cannot be compared directly. Nevertheless,  the overall picture drawn by the present study is not very dissimilar from that of Ref.~\cite{quaglioni07} or Ref.~\cite{gazit06}.  In particular, although the inclusion of the three-nucleon force and  evolved dipole operator produces a seemingly improved agreement with experiment, the considerable scatter of the experimental data  in the peak region continues to prevent a definitive conclusion concerning the quality of the interactions used. [Note that in Fig.~\ref{fig:csexpt} we estimated the total cross section from the $^4$He$(\gamma,n)$ measurements of Ref.~\cite{nilsson07} by assuming $\sigma_\gamma(\omega)\approx2\sigma_{\gamma,n}(\omega)$, and from the $^4$He$(\gamma,p)^3$H of Ref.~\cite{Raut2012} by assuming $\sigma_\gamma(\omega)\approx\sigma_{\gamma,p}(\omega)+\sigma_{\gamma,p}(\omega+0.5$ MeV).

\section{Conclusion}

We have, for the first time, SRG evolved the dipole operator in the two-body space and computed the total strength of the dipole transition, electric dipole polarizability and the total photoabsorption cross section of $^4$He. Since the dipole operator acts primarily at long range, we see little change in the convergence properties of these observables over using the bare operator.

For all three observables, there is a significant reduction of the dependence on the SRG evolution parameter when evolving the dipole operator in the two-body space. Generally, this reduction is on the order of the effect of the including the three nucleon force. So although the reduction  is relatively small in magnitude, its effects are not negligible. Any residual dependence on $\lambda$ in our calculations is due to the induced three- and four-body terms that we do not take into account.  Based on our experience with calculations of energies and radii, these higher order contributions should be smaller than the two-body contributions to the evolution.%

Future work will include evolving the dipole operator, and other nonscalar operators, in the three-body space. This will allow us to investigate the three- and four-body contribution to the evolution of these operators in the $A=4$ system. We also plan to extend these calculations to heavier systems (e.g., up to $A=12$), where it is advantageous to work with single-particle Slater determinant basis states. We will do this by transforming our two-, and eventually, three-body nonscalar operators,  presently in a translationally invariant Jacobi-coordinate basis, into matrix elements over Slater determinate basis states.

\acknowledgements
This work was performed in part under the auspices of the U.S. Department of Energy by Lawrence Livermore National Laboratory (LLNL) under Contract DE-AC52-07NA27344. This material is based upon work supported by the U.S. Department of Energy, Office of Science, Office of Nuclear Physics, under Award Numbers DE-FG02-96ER40985 and DE-FC02-07ER41457 as well as under Work Proposal Number SCW1158. Additional support came from the Natural Sciences and Engineering Research Council of Canada (NSERC) Grant Number 401945-2011. TRIUMF receives funding via a contribution through the Canadian National Research Council. Computing support came from the LLNL institutional Computing Grand Challenge program. Additional resources came from the Computational Science Research Center and Department of Physics at San Diego State University.

%This work was performed under the auspices of the U.S. Department of Energy by Lawrence Livermore National Laboratory under Contract DE-AC52-07NA27344. 
%This material is based upon work supported by the US Department of Energy, Office of Science, Office of Nuclear Physics, under Award No. DE-FG02-96ER40985 and DE-FC02-07ER41457.
%%U.S. Department of Energy grants DE-FG02-96ER40985 and DE-FC02-07ER41457 and 
%Additional support came from U.S. DOE/SC/NP (work proposal SCW1158) and The Natural Sciences and Engineering Research Council of Canada (NSERC) Grant No. 401945-2011. TRIUMF receives funding via a contribution through the National Research Council Canada. Computing support came from the LLNL institutional Computing Grand Challenge program. Additional resources came from the Computational Science Research Center and Department of Physics at San Diego State University.
% 
\bibliographystyle{apsrev}
%\bibliography{references}

\begin{thebibliography}{67}
\expandafter\ifx\csname natexlab\endcsname\relax\def\natexlab#1{#1}\fi
\expandafter\ifx\csname bibnamefont\endcsname\relax
  \def\bibnamefont#1{#1}\fi
\expandafter\ifx\csname bibfnamefont\endcsname\relax
  \def\bibfnamefont#1{#1}\fi
\expandafter\ifx\csname citenamefont\endcsname\relax
  \def\citenamefont#1{#1}\fi
\expandafter\ifx\csname url\endcsname\relax
  \def\url#1{\texttt{#1}}\fi
\expandafter\ifx\csname urlprefix\endcsname\relax\def\urlprefix{URL }\fi
\providecommand{\bibinfo}[2]{#2}
\providecommand{\eprint}[2][]{\url{#2}}

\bibitem[{\citenamefont{Okubo}(1954)}]{okubo54}
\bibinfo{author}{\bibfnamefont{S.}~\bibnamefont{Okubo}},
  \bibinfo{journal}{Prog. Theo. Phys.} \textbf{\bibinfo{volume}{12}},
  \bibinfo{pages}{603} (\bibinfo{year}{1954}).

\bibitem[{\citenamefont{Zheng et~al.}(1993)\citenamefont{Zheng, Barrett, Jaqua,
  Vary, and McCarthy}}]{zheng93}
\bibinfo{author}{\bibfnamefont{D.~C.} \bibnamefont{Zheng}},
  \bibinfo{author}{\bibfnamefont{B.~R.} \bibnamefont{Barrett}},
  \bibinfo{author}{\bibfnamefont{L.}~\bibnamefont{Jaqua}},
  \bibinfo{author}{\bibfnamefont{J.~P.} \bibnamefont{Vary}}, \bibnamefont{and}
  \bibinfo{author}{\bibfnamefont{R.~J.} \bibnamefont{McCarthy}},
  \bibinfo{journal}{Phys. Rev. C.} \textbf{\bibinfo{volume}{48}},
  \bibinfo{pages}{1083} (\bibinfo{year}{1993}).

\bibitem[{\citenamefont{Stetcu et~al.}(2007)}]{stetcu07}
\bibinfo{author}{\bibfnamefont{I.}~\bibnamefont{Stetcu}} \bibnamefont{et~al.},
  \bibinfo{journal}{Nucl. Phys. A.} \textbf{\bibinfo{volume}{785}},
  \bibinfo{pages}{307} (\bibinfo{year}{2007}).

\bibitem[{\citenamefont{Nogga et~al.}(2006)\citenamefont{Nogga, Navr\'{a}til,
  Barrett, and Vary}}]{nogga06}
\bibinfo{author}{\bibfnamefont{A.}~\bibnamefont{Nogga}},
  \bibinfo{author}{\bibfnamefont{P.}~\bibnamefont{Navr\'{a}til}},
  \bibinfo{author}{\bibfnamefont{B.~R.} \bibnamefont{Barrett}},
  \bibnamefont{and} \bibinfo{author}{\bibfnamefont{J.~P.} \bibnamefont{Vary}},
  \bibinfo{journal}{Phys. Rev. C} \textbf{\bibinfo{volume}{73}},
  \bibinfo{pages}{064002} (\bibinfo{year}{2006}).

\bibitem[{\citenamefont{Navr\'{a}til
  et~al.}(2000{\natexlab{a}})\citenamefont{Navr\'{a}til, Vary, and
  Barrett}}]{navratil00a}
\bibinfo{author}{\bibfnamefont{P.}~\bibnamefont{Navr\'{a}til}},
  \bibinfo{author}{\bibfnamefont{J.~P.} \bibnamefont{Vary}}, \bibnamefont{and}
  \bibinfo{author}{\bibfnamefont{B.~R.} \bibnamefont{Barrett}},
  \bibinfo{journal}{Phys. Rev. C} \textbf{\bibinfo{volume}{62}},
  \bibinfo{pages}{054311} (\bibinfo{year}{2000}{\natexlab{a}}).

\bibitem[{\citenamefont{Barnea et~al.}(2000)\citenamefont{Barnea, Leidemann,
  and Orlandini}}]{Barnea2000}
\bibinfo{author}{\bibfnamefont{N.}~\bibnamefont{Barnea}},
  \bibinfo{author}{\bibfnamefont{W.}~\bibnamefont{Leidemann}},
  \bibnamefont{and}
  \bibinfo{author}{\bibfnamefont{G.}~\bibnamefont{Orlandini}},
  \bibinfo{journal}{Phys. Rev. C} \textbf{\bibinfo{volume}{61}},
  \bibinfo{pages}{054001} (\bibinfo{year}{2000}).

\bibitem[{\citenamefont{Barnea et~al.}(2001)\citenamefont{Barnea, Leidemann,
  and Orlandini}}]{Barnea2001}
\bibinfo{author}{\bibfnamefont{N.}~\bibnamefont{Barnea}},
  \bibinfo{author}{\bibfnamefont{W.}~\bibnamefont{Leidemann}},
  \bibnamefont{and}
  \bibinfo{author}{\bibfnamefont{G.}~\bibnamefont{Orlandini}},
  \bibinfo{journal}{Nucl. Phys. A} \textbf{\bibinfo{volume}{693}},
  \bibinfo{pages}{565 } (\bibinfo{year}{2001}).

\bibitem[{\citenamefont{Jurgenson et~al.}(2011)\citenamefont{Jurgenson,
  Navr\'{a}til, and Furnstahl}}]{jurgenson11}
\bibinfo{author}{\bibfnamefont{E.~D.} \bibnamefont{Jurgenson}},
  \bibinfo{author}{\bibfnamefont{P.}~\bibnamefont{Navr\'{a}til}},
  \bibnamefont{and} \bibinfo{author}{\bibfnamefont{R.~J.}
  \bibnamefont{Furnstahl}}, \bibinfo{journal}{Phys. Rev. C}
  \textbf{\bibinfo{volume}{83}}, \bibinfo{pages}{034301}
  (\bibinfo{year}{2011}).

\bibitem[{\citenamefont{Roth et~al.}(2011)\citenamefont{Roth, Langhammer,
  Calci, Binder, and Navr\'{a}til}}]{roth11a}
\bibinfo{author}{\bibfnamefont{R.}~\bibnamefont{Roth}},
  \bibinfo{author}{\bibfnamefont{J.}~\bibnamefont{Langhammer}},
  \bibinfo{author}{\bibfnamefont{A.}~\bibnamefont{Calci}},
  \bibinfo{author}{\bibfnamefont{S.}~\bibnamefont{Binder}}, \bibnamefont{and}
  \bibinfo{author}{\bibfnamefont{P.}~\bibnamefont{Navr\'{a}til}},
  \bibinfo{journal}{Phys. Rev. Lett.} \textbf{\bibinfo{volume}{107}},
  \bibinfo{pages}{072501} (\bibinfo{year}{2011}).

\bibitem[{\citenamefont{Stetcu and Rotureau}(2013)}]{stetcu12}
\bibinfo{author}{\bibfnamefont{I.}~\bibnamefont{Stetcu}} \bibnamefont{and}
  \bibinfo{author}{\bibfnamefont{J.}~\bibnamefont{Rotureau}},
  \bibinfo{journal}{Prog. Part. Nucl. Phys.} \textbf{\bibinfo{volume}{69}},
  \bibinfo{pages}{182} (\bibinfo{year}{2013}).

\bibitem[{\citenamefont{Furnstahl and Hebeler}(2013)}]{furnstahl13a}
\bibinfo{author}{\bibfnamefont{R.~J.} \bibnamefont{Furnstahl}}
  \bibnamefont{and} \bibinfo{author}{\bibfnamefont{K.}~\bibnamefont{Hebeler}},
  \bibinfo{journal}{Rep. Prog. Phys.} \textbf{\bibinfo{volume}{76}},
  \bibinfo{pages}{126301} (\bibinfo{year}{2013}).

\bibitem[{\citenamefont{Bogner et~al.}(2007)\citenamefont{Bogner, Furnstahl,
  and Perry}}]{bogner07}
\bibinfo{author}{\bibfnamefont{S.~K.} \bibnamefont{Bogner}},
  \bibinfo{author}{\bibfnamefont{R.~J.} \bibnamefont{Furnstahl}},
  \bibnamefont{and} \bibinfo{author}{\bibfnamefont{R.~J.} \bibnamefont{Perry}},
  \bibinfo{journal}{Phys. Rev. C} \textbf{\bibinfo{volume}{75}},
  \bibinfo{pages}{061001} (\bibinfo{year}{2007}).

\bibitem[{\citenamefont{Jurgenson et~al.}(2009)\citenamefont{Jurgenson,
  Navr\'{a}til, and Furnstahl}}]{jurgenson09a}
\bibinfo{author}{\bibfnamefont{E.~D.} \bibnamefont{Jurgenson}},
  \bibinfo{author}{\bibfnamefont{P.}~\bibnamefont{Navr\'{a}til}},
  \bibnamefont{and} \bibinfo{author}{\bibfnamefont{R.~J.}
  \bibnamefont{Furnstahl}}, \bibinfo{journal}{Phys. Rev. Lett.}
  \textbf{\bibinfo{volume}{103}}, \bibinfo{pages}{082501}
  (\bibinfo{year}{2009}).

\bibitem[{\citenamefont{Roth et~al.}(2012)\citenamefont{Roth, Binder, Vobig,
  Calci, Langhammer, and Navr\'{a}til}}]{roth12a}
\bibinfo{author}{\bibfnamefont{R.}~\bibnamefont{Roth}},
  \bibinfo{author}{\bibfnamefont{S.}~\bibnamefont{Binder}},
  \bibinfo{author}{\bibfnamefont{K.}~\bibnamefont{Vobig}},
  \bibinfo{author}{\bibfnamefont{A.}~\bibnamefont{Calci}},
  \bibinfo{author}{\bibfnamefont{J.}~\bibnamefont{Langhammer}},
  \bibnamefont{and}
  \bibinfo{author}{\bibfnamefont{P.}~\bibnamefont{Navr\'{a}til}},
  \bibinfo{journal}{Phys. Rev. Lett.} \textbf{\bibinfo{volume}{109}},
  \bibinfo{pages}{052501} (\bibinfo{year}{2012}).

\bibitem[{\citenamefont{Dicaire
  et~al.}(2014{\natexlab{a}})\citenamefont{Dicaire, Omand, and
  Navr\'atil}}]{Dicaire2014}
\bibinfo{author}{\bibfnamefont{N.~M.} \bibnamefont{Dicaire}},
  \bibinfo{author}{\bibfnamefont{C.}~\bibnamefont{Omand}}, \bibnamefont{and}
  \bibinfo{author}{\bibfnamefont{P.}~\bibnamefont{Navr\'atil}},
  \bibinfo{journal}{Phys. Rev. C} \textbf{\bibinfo{volume}{90}},
  \bibinfo{pages}{034302} (\bibinfo{year}{2014}{\natexlab{a}}).

\bibitem[{\citenamefont{Wendt}(2013)}]{Wendt2013}
\bibinfo{author}{\bibfnamefont{K.~A.} \bibnamefont{Wendt}},
  \bibinfo{journal}{Phys. Rev. C} \textbf{\bibinfo{volume}{87}},
  \bibinfo{pages}{061001(R)} (\bibinfo{year}{2013}).

\bibitem[{\citenamefont{Tsukiyama et~al.}(2011)\citenamefont{Tsukiyama, Bogner,
  and Schwenk}}]{tsukiyamat11a}
\bibinfo{author}{\bibfnamefont{K.}~\bibnamefont{Tsukiyama}},
  \bibinfo{author}{\bibfnamefont{S.~K.} \bibnamefont{Bogner}},
  \bibnamefont{and} \bibinfo{author}{\bibfnamefont{A.}~\bibnamefont{Schwenk}},
  \bibinfo{journal}{Phys. Rev. Lett.} \textbf{\bibinfo{volume}{106}},
  \bibinfo{pages}{222502} (\bibinfo{year}{2011}).

\bibitem[{\citenamefont{Hergert et~al.}(2013)}]{hergert13a}
\bibinfo{author}{\bibfnamefont{H.}~\bibnamefont{Hergert}} \bibnamefont{et~al.},
  \bibinfo{journal}{Phys. Rev. C} \textbf{\bibinfo{volume}{87}},
  \bibinfo{pages}{034307} (\bibinfo{year}{2013}).

\bibitem[{\citenamefont{Hupin et~al.}(2014)\citenamefont{Hupin, Quaglioni, and
  Navr\'{a}til}}]{hupin2014}
\bibinfo{author}{\bibfnamefont{G.}~\bibnamefont{Hupin}},
  \bibinfo{author}{\bibfnamefont{S.}~\bibnamefont{Quaglioni}},
  \bibnamefont{and}
  \bibinfo{author}{\bibfnamefont{P.}~\bibnamefont{Navr\'{a}til}},
  \bibinfo{journal}{arXiv:1409.0892v1}  (\bibinfo{year}{2014}).

\bibitem[{\citenamefont{Hupin et~al.}(2013)\citenamefont{Hupin, Langhammer,
  Navr\'{a}til, Quaglioni, Calci, and Roth}}]{Hupin2013}
\bibinfo{author}{\bibfnamefont{G.}~\bibnamefont{Hupin}},
  \bibinfo{author}{\bibfnamefont{J.}~\bibnamefont{Langhammer}},
  \bibinfo{author}{\bibfnamefont{P.}~\bibnamefont{Navr\'{a}til}},
  \bibinfo{author}{\bibfnamefont{S.}~\bibnamefont{Quaglioni}},
  \bibinfo{author}{\bibfnamefont{A.}~\bibnamefont{Calci}}, \bibnamefont{and}
  \bibinfo{author}{\bibfnamefont{R.}~\bibnamefont{Roth}},
  \bibinfo{journal}{Phys. Rev. C} \textbf{\bibinfo{volume}{88}},
  \bibinfo{pages}{054622} (\bibinfo{year}{2013}).

\bibitem[{\citenamefont{Langhammer et~al.}(2015)\citenamefont{Langhammer,
  Navr\'{a}til, Quaglioni, Hupin, Calci, and Roth}}]{Langhammer2015}
\bibinfo{author}{\bibfnamefont{J.}~\bibnamefont{Langhammer}},
  \bibinfo{author}{\bibfnamefont{P.}~\bibnamefont{Navr\'{a}til}},
  \bibinfo{author}{\bibfnamefont{S.}~\bibnamefont{Quaglioni}},
  \bibinfo{author}{\bibfnamefont{G.}~\bibnamefont{Hupin}},
  \bibinfo{author}{\bibfnamefont{A.}~\bibnamefont{Calci}}, \bibnamefont{and}
  \bibinfo{author}{\bibfnamefont{R.}~\bibnamefont{Roth}},
  \bibinfo{journal}{Phys. Rev. C} \textbf{\bibinfo{volume}{91}},
  \bibinfo{pages}{021301(R)} (\bibinfo{year}{2015}).

\bibitem[{\citenamefont{Suzuki and Lee}(1980)}]{suzuki80}
\bibinfo{author}{\bibfnamefont{K.}~\bibnamefont{Suzuki}} \bibnamefont{and}
  \bibinfo{author}{\bibfnamefont{S.~Y.} \bibnamefont{Lee}},
  \bibinfo{journal}{Prog. Theor. Phys} \textbf{\bibinfo{volume}{64}},
  \bibinfo{pages}{2091} (\bibinfo{year}{1980}).

\bibitem[{\citenamefont{Suzuki}(1982)}]{suzuki82}
\bibinfo{author}{\bibfnamefont{K.}~\bibnamefont{Suzuki}},
  \bibinfo{journal}{Prog. Theor. Phys} \textbf{\bibinfo{volume}{68}},
  \bibinfo{pages}{1999} (\bibinfo{year}{1982}).

\bibitem[{\citenamefont{Stetcu et~al.}(2005)\citenamefont{Stetcu, Barrett,
  Navr\'{a}til, and Vary}}]{stetcu05}
\bibinfo{author}{\bibfnamefont{I.}~\bibnamefont{Stetcu}},
  \bibinfo{author}{\bibfnamefont{B.~R.} \bibnamefont{Barrett}},
  \bibinfo{author}{\bibfnamefont{P.}~\bibnamefont{Navr\'{a}til}},
  \bibnamefont{and} \bibinfo{author}{\bibfnamefont{J.~P.} \bibnamefont{Vary}},
  \bibinfo{journal}{Phys. Rev. C} \textbf{\bibinfo{volume}{71}},
  \bibinfo{pages}{044325} (\bibinfo{year}{2005}).

\bibitem[{\citenamefont{Anderson et~al.}(2010)\citenamefont{Anderson, Bogner,
  Furnstahl, and Perry}}]{anderson10}
\bibinfo{author}{\bibfnamefont{E.~R.} \bibnamefont{Anderson}},
  \bibinfo{author}{\bibfnamefont{S.~K.} \bibnamefont{Bogner}},
  \bibinfo{author}{\bibfnamefont{R.~J.} \bibnamefont{Furnstahl}},
  \bibnamefont{and} \bibinfo{author}{\bibfnamefont{R.~J.} \bibnamefont{Perry}},
  \bibinfo{journal}{Phys. Rev. C} \textbf{\bibinfo{volume}{82}},
  \bibinfo{pages}{054001} (\bibinfo{year}{2010}).

\bibitem[{\citenamefont{Schuster et~al.}(2014)\citenamefont{Schuster,
  Quaglioni, Johnson, Jurgenson, and Navr\'{a}til}}]{schuster14a}
\bibinfo{author}{\bibfnamefont{M.~D.} \bibnamefont{Schuster}},
  \bibinfo{author}{\bibfnamefont{S.}~\bibnamefont{Quaglioni}},
  \bibinfo{author}{\bibfnamefont{C.~W.} \bibnamefont{Johnson}},
  \bibinfo{author}{\bibfnamefont{E.~D.} \bibnamefont{Jurgenson}},
  \bibnamefont{and}
  \bibinfo{author}{\bibfnamefont{P.}~\bibnamefont{Navr\'{a}til}},
  \bibinfo{journal}{Phys. Rev. C} \textbf{\bibinfo{volume}{90}},
  \bibinfo{pages}{011301(R)} (\bibinfo{year}{2014}).

\bibitem[{\citenamefont{Epelbaum et~al.}(2009)\citenamefont{Epelbaum, Hammer,
  and Mei{\ss}ner}}]{epelbaum09a}
\bibinfo{author}{\bibfnamefont{E.}~\bibnamefont{Epelbaum}},
  \bibinfo{author}{\bibfnamefont{H.-W.} \bibnamefont{Hammer}},
  \bibnamefont{and} \bibinfo{author}{\bibfnamefont{U.-G.}
  \bibnamefont{Mei{\ss}ner}}, \bibinfo{journal}{Rev. Mod. Phys.}
  \textbf{\bibinfo{volume}{81}}, \bibinfo{pages}{1773} (\bibinfo{year}{2009}).

\bibitem[{\citenamefont{Machleidt and Entem}(2011)}]{machleidt11a}
\bibinfo{author}{\bibfnamefont{R.}~\bibnamefont{Machleidt}} \bibnamefont{and}
  \bibinfo{author}{\bibfnamefont{D.~R.} \bibnamefont{Entem}},
  \bibinfo{journal}{Phys. Rep.} \textbf{\bibinfo{volume}{503}},
  \bibinfo{pages}{1} (\bibinfo{year}{2011}).

\bibitem[{\citenamefont{Navr\'{a}til
  et~al.}(2000{\natexlab{b}})\citenamefont{Navr\'{a}til, Kamuntavi{\v{c}}ius,
  and Barrett}}]{navratil00b}
\bibinfo{author}{\bibfnamefont{P.}~\bibnamefont{Navr\'{a}til}},
  \bibinfo{author}{\bibfnamefont{G.~P.} \bibnamefont{Kamuntavi{\v{c}}ius}},
  \bibnamefont{and} \bibinfo{author}{\bibfnamefont{B.~R.}
  \bibnamefont{Barrett}}, \bibinfo{journal}{Phys. Rev. C}
  \textbf{\bibinfo{volume}{61}}, \bibinfo{pages}{044001}
  (\bibinfo{year}{2000}{\natexlab{b}}).

\bibitem[{\citenamefont{Efros et~al.}(1994)\citenamefont{Efros, Leidemann, and
  Orlandini}}]{efros94a}
\bibinfo{author}{\bibfnamefont{V.~D.} \bibnamefont{Efros}},
  \bibinfo{author}{\bibfnamefont{W.}~\bibnamefont{Leidemann}},
  \bibnamefont{and}
  \bibinfo{author}{\bibfnamefont{G.}~\bibnamefont{Orlandini}},
  \bibinfo{journal}{Phys. Lett. B} \textbf{\bibinfo{volume}{338}},
  \bibinfo{pages}{130} (\bibinfo{year}{1994}).

\bibitem[{\citenamefont{Efros et~al.}(2007)\citenamefont{Efros, Leidemann,
  Orlandini, and Barnea}}]{efros07a}
\bibinfo{author}{\bibfnamefont{V.~D.} \bibnamefont{Efros}},
  \bibinfo{author}{\bibfnamefont{W.}~\bibnamefont{Leidemann}},
  \bibinfo{author}{\bibfnamefont{G.}~\bibnamefont{Orlandini}},
  \bibnamefont{and} \bibinfo{author}{\bibfnamefont{N.}~\bibnamefont{Barnea}},
  \bibinfo{journal}{J. Phys. G} \textbf{\bibinfo{volume}{34}},
  \bibinfo{pages}{R459} (\bibinfo{year}{2007}).

\bibitem[{\citenamefont{Podolsky}(1928)}]{Podolsky1928}
\bibinfo{author}{\bibfnamefont{B.}~\bibnamefont{Podolsky}},
  \bibinfo{journal}{Proc. Natl. Acad. Sci. U.S.A.}
  \textbf{\bibinfo{volume}{14}}, \bibinfo{pages}{253} (\bibinfo{year}{1928}).

\bibitem[{\citenamefont{Quaglioni and Navr\'{a}til}(2007)}]{quaglioni07}
\bibinfo{author}{\bibfnamefont{S.}~\bibnamefont{Quaglioni}} \bibnamefont{and}
  \bibinfo{author}{\bibfnamefont{P.}~\bibnamefont{Navr\'{a}til}},
  \bibinfo{journal}{Phys. Lett. B} \textbf{\bibinfo{volume}{652}},
  \bibinfo{pages}{370} (\bibinfo{year}{2007}).

\bibitem[{\citenamefont{Stetcu et~al.}(2009)\citenamefont{Stetcu, Quaglioni,
  Friar, Hayes, and Navr\'{a}til}}]{stetcu09a}
\bibinfo{author}{\bibfnamefont{I.}~\bibnamefont{Stetcu}},
  \bibinfo{author}{\bibfnamefont{S.}~\bibnamefont{Quaglioni}},
  \bibinfo{author}{\bibfnamefont{J.~L.} \bibnamefont{Friar}},
  \bibinfo{author}{\bibfnamefont{A.~C.} \bibnamefont{Hayes}}, \bibnamefont{and}
  \bibinfo{author}{\bibfnamefont{P.}~\bibnamefont{Navr\'{a}til}},
  \bibinfo{journal}{Phys. Rev. C} \textbf{\bibinfo{volume}{79}},
  \bibinfo{pages}{064001} (\bibinfo{year}{2009}).

\bibitem[{\citenamefont{Navr\'atil and Ormand}(2002)}]{Navratil2002}
\bibinfo{author}{\bibfnamefont{P.}~\bibnamefont{Navr\'atil}} \bibnamefont{and}
  \bibinfo{author}{\bibfnamefont{W.~E.} \bibnamefont{Ormand}},
  \bibinfo{journal}{Phys. Rev. Lett.} \textbf{\bibinfo{volume}{88}},
  \bibinfo{pages}{152502} (\bibinfo{year}{2002}).

\bibitem[{\citenamefont{Navr\'{a}til and Ormand}(2003)}]{navratil03a}
\bibinfo{author}{\bibfnamefont{P.}~\bibnamefont{Navr\'{a}til}}
  \bibnamefont{and} \bibinfo{author}{\bibfnamefont{W.~E.}
  \bibnamefont{Ormand}}, \bibinfo{journal}{Phys. Rev. C}
  \textbf{\bibinfo{volume}{68}}, \bibinfo{pages}{034305}
  (\bibinfo{year}{2003}).

\bibitem[{\citenamefont{Lanczos}(1950)}]{lanczos}
\bibinfo{author}{\bibfnamefont{C.}~\bibnamefont{Lanczos}}, \bibinfo{journal}{J.
  Res. Natl. Bur. Stand.} \textbf{\bibinfo{volume}{45}}, \bibinfo{pages}{255}
  (\bibinfo{year}{1950}).

\bibitem[{\citenamefont{Haydock}(1974)}]{haydock74}
\bibinfo{author}{\bibfnamefont{R.}~\bibnamefont{Haydock}}, \bibinfo{journal}{J.
  Phys. A: Math. Nucl. Gen.} \textbf{\bibinfo{volume}{7}},
  \bibinfo{pages}{2120} (\bibinfo{year}{1974}).

\bibitem[{\citenamefont{Haxton et~al.}(2005)\citenamefont{Haxton, Nollett, and
  Zurek}}]{haxton05a}
\bibinfo{author}{\bibfnamefont{W.~C.} \bibnamefont{Haxton}},
  \bibinfo{author}{\bibfnamefont{K.~M.} \bibnamefont{Nollett}},
  \bibnamefont{and} \bibinfo{author}{\bibfnamefont{K.~M.} \bibnamefont{Zurek}},
  \bibinfo{journal}{Phys. Rev. C} \textbf{\bibinfo{volume}{72}},
  \bibinfo{pages}{065501} (\bibinfo{year}{2005}).

\bibitem[{\citenamefont{Bogner et~al.}(2010)\citenamefont{Bogner, Furnstahl,
  and Schwenk}}]{bogner10}
\bibinfo{author}{\bibfnamefont{S.~K.} \bibnamefont{Bogner}},
  \bibinfo{author}{\bibfnamefont{R.~J.} \bibnamefont{Furnstahl}},
  \bibnamefont{and} \bibinfo{author}{\bibfnamefont{A.}~\bibnamefont{Schwenk}},
  \bibinfo{journal}{Prog. Part. Nucl. Phys.} \textbf{\bibinfo{volume}{65}},
  \bibinfo{pages}{94} (\bibinfo{year}{2010}).

\bibitem[{\citenamefont{Wegner}(1994)}]{wegner94}
\bibinfo{author}{\bibfnamefont{F.}~\bibnamefont{Wegner}},
  \bibinfo{journal}{Ann. Physik} \textbf{\bibinfo{volume}{3}},
  \bibinfo{pages}{77} (\bibinfo{year}{1994}).

\bibitem[{\citenamefont{Li et~al.}(2011)\citenamefont{Li, Anderson, and
  Furnstahl}}]{li11}
\bibinfo{author}{\bibfnamefont{W.}~\bibnamefont{Li}},
  \bibinfo{author}{\bibfnamefont{E.~R.} \bibnamefont{Anderson}},
  \bibnamefont{and} \bibinfo{author}{\bibfnamefont{R.~J.}
  \bibnamefont{Furnstahl}}, \bibinfo{journal}{Phys. Rev. C}
  \textbf{\bibinfo{volume}{84}}, \bibinfo{pages}{054002}
  (\bibinfo{year}{2011}).

\bibitem[{\citenamefont{Dicaire
  et~al.}(2014{\natexlab{b}})\citenamefont{Dicaire, Omand, and
  Navr\'{a}til}}]{dicaire14a}
\bibinfo{author}{\bibfnamefont{N.~M.} \bibnamefont{Dicaire}},
  \bibinfo{author}{\bibfnamefont{C.}~\bibnamefont{Omand}}, \bibnamefont{and}
  \bibinfo{author}{\bibfnamefont{P.}~\bibnamefont{Navr\'{a}til}},
  \bibinfo{journal}{Phys. Rev. C} \textbf{\bibinfo{volume}{90}},
  \bibinfo{pages}{034302} (\bibinfo{year}{2014}{\natexlab{b}}).

\bibitem[{\citenamefont{Levinger and Bethe}(1950)}]{levinger50}
\bibinfo{author}{\bibfnamefont{J.~S.} \bibnamefont{Levinger}} \bibnamefont{and}
  \bibinfo{author}{\bibfnamefont{H.~A.} \bibnamefont{Bethe}},
  \bibinfo{journal}{Phys. Rev.} \textbf{\bibinfo{volume}{78}},
  \bibinfo{pages}{115} (\bibinfo{year}{1950}).

\bibitem[{\citenamefont{Efros et~al.}(1999)\citenamefont{Efros, Leidemann, and
  Orlandini}}]{efros99a}
\bibinfo{author}{\bibfnamefont{V.~D.} \bibnamefont{Efros}},
  \bibinfo{author}{\bibfnamefont{W.}~\bibnamefont{Leidemann}},
  \bibnamefont{and}
  \bibinfo{author}{\bibfnamefont{G.}~\bibnamefont{Orlandini}},
  \bibinfo{journal}{Few-body Syst.} \textbf{\bibinfo{volume}{26}},
  \bibinfo{pages}{251} (\bibinfo{year}{1999}).

\bibitem[{\citenamefont{Andreasi et~al.}(2005)\citenamefont{Andreasi,
  Leidemann, Rei\ss, and Schwamb}}]{andreasi05a}
\bibinfo{author}{\bibfnamefont{D.}~\bibnamefont{Andreasi}},
  \bibinfo{author}{\bibfnamefont{W.}~\bibnamefont{Leidemann}},
  \bibinfo{author}{\bibfnamefont{C.}~\bibnamefont{Rei\ss}}, \bibnamefont{and}
  \bibinfo{author}{\bibfnamefont{M.}~\bibnamefont{Schwamb}},
  \bibinfo{journal}{Eur. Phys. J. A} \textbf{\bibinfo{volume}{24}},
  \bibinfo{pages}{361} (\bibinfo{year}{2005}).

\bibitem[{\citenamefont{Marchisio et~al.}(2003)\citenamefont{Marchisio, Barnea,
  Leidemann, and Orlandini}}]{marchisio03}
\bibinfo{author}{\bibfnamefont{M.~A.} \bibnamefont{Marchisio}},
  \bibinfo{author}{\bibfnamefont{N.}~\bibnamefont{Barnea}},
  \bibinfo{author}{\bibfnamefont{W.}~\bibnamefont{Leidemann}},
  \bibnamefont{and}
  \bibinfo{author}{\bibfnamefont{G.}~\bibnamefont{Orlandini}},
  \bibinfo{journal}{Few-Body Syst.} \textbf{\bibinfo{volume}{33}},
  \bibinfo{pages}{259} (\bibinfo{year}{2003}).

\bibitem[{\citenamefont{Entem and Machleidt}(2003)}]{entem03a}
\bibinfo{author}{\bibfnamefont{D.~R.} \bibnamefont{Entem}} \bibnamefont{and}
  \bibinfo{author}{\bibfnamefont{R.}~\bibnamefont{Machleidt}},
  \bibinfo{journal}{Phys. Rev. C} \textbf{\bibinfo{volume}{68}},
  \bibinfo{pages}{041001} (\bibinfo{year}{2003}).

\bibitem[{\citenamefont{Navr\'{a}til}(2007)}]{navratil07a}
\bibinfo{author}{\bibfnamefont{P.}~\bibnamefont{Navr\'{a}til}},
  \bibinfo{journal}{Few Body Syst.} \textbf{\bibinfo{volume}{41}},
  \bibinfo{pages}{117} (\bibinfo{year}{2007}).

\bibitem[{\citenamefont{Gazit et~al.}(2009)\citenamefont{Gazit, Quaglioni, and
  Navr\'{a}til}}]{gazit09a}
\bibinfo{author}{\bibfnamefont{D.}~\bibnamefont{Gazit}},
  \bibinfo{author}{\bibfnamefont{S.}~\bibnamefont{Quaglioni}},
  \bibnamefont{and}
  \bibinfo{author}{\bibfnamefont{P.}~\bibnamefont{Navr\'{a}til}},
  \bibinfo{journal}{Phys. Rev. Lett.} \textbf{\bibinfo{volume}{103}},
  \bibinfo{pages}{102502} (\bibinfo{year}{2009}).

\bibitem[{\citenamefont{Roth et~al.}(2014)\citenamefont{Roth, Calci,
  Langhammer, and Binder}}]{roth14a}
\bibinfo{author}{\bibfnamefont{R.}~\bibnamefont{Roth}},
  \bibinfo{author}{\bibfnamefont{A.}~\bibnamefont{Calci}},
  \bibinfo{author}{\bibfnamefont{J.}~\bibnamefont{Langhammer}},
  \bibnamefont{and} \bibinfo{author}{\bibfnamefont{S.}~\bibnamefont{Binder}},
  \bibinfo{journal}{Phys. Rev. C} \textbf{\bibinfo{volume}{90}},
  \bibinfo{pages}{024325} (\bibinfo{year}{2014}).

\bibitem[{\citenamefont{Anderson et~al.}(2008)\citenamefont{Anderson, Bogner,
  Furnstahl, Jurgenson, Perry, and Schwenk}}]{anderson08a}
\bibinfo{author}{\bibfnamefont{E.}~\bibnamefont{Anderson}},
  \bibinfo{author}{\bibfnamefont{S.~K.} \bibnamefont{Bogner}},
  \bibinfo{author}{\bibfnamefont{R.~J.} \bibnamefont{Furnstahl}},
  \bibinfo{author}{\bibfnamefont{E.~D.} \bibnamefont{Jurgenson}},
  \bibinfo{author}{\bibfnamefont{R.~J.} \bibnamefont{Perry}}, \bibnamefont{and}
  \bibinfo{author}{\bibfnamefont{A.}~\bibnamefont{Schwenk}},
  \bibinfo{journal}{Phys. Rev. C} \textbf{\bibinfo{volume}{77}},
  \bibinfo{pages}{037001} (\bibinfo{year}{2008}).

\bibitem[{\citenamefont{Coon and Kruse}(2014)}]{CK13}
\bibinfo{author}{\bibfnamefont{S.~A.} \bibnamefont{Coon}} \bibnamefont{and}
  \bibinfo{author}{\bibfnamefont{M.~K.~G.} \bibnamefont{Kruse}}, in
  \emph{\bibinfo{booktitle}{Proceedings of International Workshop on Nuclear
  Theory in the Supercomputing Era (NTSE-2013), Ames, Iowa, 2013}}, edited by
  \bibinfo{editor}{\bibfnamefont{A.}~\bibnamefont{Shirokov}} \bibnamefont{and}
  \bibinfo{editor}{\bibfnamefont{A.}~\bibnamefont{Mazur}}
  (\bibinfo{address}{Pacific National University, Khabarovsk},
  \bibinfo{year}{2014}), pp. \bibinfo{pages}{314--324}.

\bibitem[{\citenamefont{Tilley et~al.}(1992)\citenamefont{Tilley, Weller, and
  Hale}}]{tilley92a}
\bibinfo{author}{\bibfnamefont{D.~R.} \bibnamefont{Tilley}},
  \bibinfo{author}{\bibfnamefont{H.~R.} \bibnamefont{Weller}},
  \bibnamefont{and} \bibinfo{author}{\bibfnamefont{G.~M.} \bibnamefont{Hale}},
  \bibinfo{journal}{Nucl. Phys. A} \textbf{\bibinfo{volume}{541}},
  \bibinfo{pages}{1} (\bibinfo{year}{1992}).

\bibitem[{rad()}]{radius}
\bibinfo{note}{The experimental value of the point-proton radius is deduced
  from the measured $^4$He charge radius, $\sqrt{\langle r_c^2\rangle}
  =1.673(1)$ fm~\cite{Borie1978}, proton charge radius, $\sqrt{\langle
  R_p^2\rangle} =0.895(18)$ fm~\cite{Sick2003}, and neutron mean-square-charge
  radius, $\langle R_n^2\rangle =0.120(5)$ fm$^2$~\cite{Kopecky1995}}.

\bibitem[{\citenamefont{Friar}(1977)}]{Friar1977}
\bibinfo{author}{\bibfnamefont{J.~L.} \bibnamefont{Friar}},
  \bibinfo{journal}{Phys. Rev. C} \textbf{\bibinfo{volume}{16}},
  \bibinfo{pages}{1540} (\bibinfo{year}{1977}).

\bibitem[{\citenamefont{Pachucki and Moro}(2007)}]{Pachucki2007}
\bibinfo{author}{\bibfnamefont{K.}~\bibnamefont{Pachucki}} \bibnamefont{and}
  \bibinfo{author}{\bibfnamefont{A.~M.} \bibnamefont{Moro}},
  \bibinfo{journal}{Phys. Rev. A} \textbf{\bibinfo{volume}{75}},
  \bibinfo{pages}{032521} (\bibinfo{year}{2007}).

\bibitem[{\citenamefont{Wells et~al.}(1992)}]{wells92}
\bibinfo{author}{\bibfnamefont{D.~P.} \bibnamefont{Wells}}
  \bibnamefont{et~al.}, \bibinfo{journal}{Phys. Rev. C}
  \textbf{\bibinfo{volume}{46}}, \bibinfo{pages}{449} (\bibinfo{year}{1992}).

\bibitem[{\citenamefont{Shima et~al.}(2005)}]{shima05}
\bibinfo{author}{\bibfnamefont{T.}~\bibnamefont{Shima}} \bibnamefont{et~al.},
  \bibinfo{journal}{Phys. Rev. C} \textbf{\bibinfo{volume}{72}},
  \bibinfo{pages}{044004} (\bibinfo{year}{2005}).

\bibitem[{\citenamefont{Nilsson et~al.}(2007)}]{nilsson07}
\bibinfo{author}{\bibfnamefont{B.}~\bibnamefont{Nilsson}} \bibnamefont{et~al.},
  \bibinfo{journal}{Phys. Rev. C} \textbf{\bibinfo{volume}{75}},
  \bibinfo{pages}{014007} (\bibinfo{year}{2007}).

\bibitem[{\citenamefont{Nakayama et~al.}(2007)}]{nakayama07}
\bibinfo{author}{\bibfnamefont{S.}~\bibnamefont{Nakayama}}
  \bibnamefont{et~al.}, \bibinfo{journal}{Phys. Rev. C}
  \textbf{\bibinfo{volume}{76}}, \bibinfo{pages}{021305(R)}
  (\bibinfo{year}{2007}).

\bibitem[{\citenamefont{Raut et~al.}(2012)\citenamefont{Raut, Tornow, Ahmed,
  Crowell, Kelley, Rusev, Stave, and Tonchev}}]{Raut2012}
\bibinfo{author}{\bibfnamefont{R.}~\bibnamefont{Raut}},
  \bibinfo{author}{\bibfnamefont{W.}~\bibnamefont{Tornow}},
  \bibinfo{author}{\bibfnamefont{M.~W.} \bibnamefont{Ahmed}},
  \bibinfo{author}{\bibfnamefont{A.~S.} \bibnamefont{Crowell}},
  \bibinfo{author}{\bibfnamefont{J.~H.} \bibnamefont{Kelley}},
  \bibinfo{author}{\bibfnamefont{G.}~\bibnamefont{Rusev}},
  \bibinfo{author}{\bibfnamefont{S.~C.} \bibnamefont{Stave}}, \bibnamefont{and}
  \bibinfo{author}{\bibfnamefont{A.~P.} \bibnamefont{Tonchev}},
  \bibinfo{journal}{Phys. Rev. Lett.} \textbf{\bibinfo{volume}{108}},
  \bibinfo{pages}{042502} (\bibinfo{year}{2012}).

\bibitem[{\citenamefont{Gazit et~al.}(2006)\citenamefont{Gazit, Bacca, Barnea,
  Leidemann, and Orlandini}}]{gazit06}
\bibinfo{author}{\bibfnamefont{D.}~\bibnamefont{Gazit}},
  \bibinfo{author}{\bibfnamefont{S.}~\bibnamefont{Bacca}},
  \bibinfo{author}{\bibfnamefont{N.}~\bibnamefont{Barnea}},
  \bibinfo{author}{\bibfnamefont{W.}~\bibnamefont{Leidemann}},
  \bibnamefont{and}
  \bibinfo{author}{\bibfnamefont{G.}~\bibnamefont{Orlandini}},
  \bibinfo{journal}{Phys. Rev. Lett.} \textbf{\bibinfo{volume}{96}},
  \bibinfo{pages}{112301} (\bibinfo{year}{2006}).

\bibitem[{\citenamefont{Bacca et~al.}(2014)\citenamefont{Bacca, Barnea, Hagen,
  Miorelli, Orlandini, and Papenbrock}}]{bacca14}
\bibinfo{author}{\bibfnamefont{S.}~\bibnamefont{Bacca}},
  \bibinfo{author}{\bibfnamefont{N.}~\bibnamefont{Barnea}},
  \bibinfo{author}{\bibfnamefont{G.}~\bibnamefont{Hagen}},
  \bibinfo{author}{\bibfnamefont{M.}~\bibnamefont{Miorelli}},
  \bibinfo{author}{\bibfnamefont{G.}~\bibnamefont{Orlandini}},
  \bibnamefont{and}
  \bibinfo{author}{\bibfnamefont{T.}~\bibnamefont{Papenbrock}},
  \bibinfo{journal}{Phys. Rev. C} \textbf{\bibinfo{volume}{90}},
  \bibinfo{pages}{064619} (\bibinfo{year}{2014}).

\bibitem[{\citenamefont{Borie and Rinker}(1978)}]{Borie1978}
\bibinfo{author}{\bibfnamefont{E.}~\bibnamefont{Borie}} \bibnamefont{and}
  \bibinfo{author}{\bibfnamefont{G.~A.} \bibnamefont{Rinker}},
  \bibinfo{journal}{Phys. Rev. A} \textbf{\bibinfo{volume}{18}},
  \bibinfo{pages}{324} (\bibinfo{year}{1978}).

\bibitem[{\citenamefont{Sick}(2003)}]{Sick2003}
\bibinfo{author}{\bibfnamefont{I.}~\bibnamefont{Sick}}, \bibinfo{journal}{Phys.
  Lett. B} \textbf{\bibinfo{volume}{576}}, \bibinfo{pages}{62 }
  (\bibinfo{year}{2003}).

\bibitem[{\citenamefont{Kopecky et~al.}(1995)\citenamefont{Kopecky, Riehs,
  Harvey, and Hill}}]{Kopecky1995}
\bibinfo{author}{\bibfnamefont{S.}~\bibnamefont{Kopecky}},
  \bibinfo{author}{\bibfnamefont{P.}~\bibnamefont{Riehs}},
  \bibinfo{author}{\bibfnamefont{J.~A.} \bibnamefont{Harvey}},
  \bibnamefont{and} \bibinfo{author}{\bibfnamefont{N.~W.} \bibnamefont{Hill}},
  \bibinfo{journal}{Phys. Rev. Lett.} \textbf{\bibinfo{volume}{74}},
  \bibinfo{pages}{2427} (\bibinfo{year}{1995}).

\end{thebibliography}

\end{document}